\documentclass[fleqn,usenatbib]{mnras}

\usepackage{newtxtext,newtxmath}

\usepackage[T1]{fontenc}
\usepackage{ae,aecompl}

\usepackage{graphicx}	
\usepackage{color}
\usepackage{amsmath}	
\usepackage{bm}
\usepackage{natbib}
\usepackage{pgffor}

\makeatletter
\def\dif{\@ifnextchar[{\@with}{\@without}}
\def\@with[#1]#2{
  \ensuremath{\frac{\foreach \x in {#1}{\mathrm{d}\x\,}}{\foreach \x in {#2}{\mathrm{d}\x\,}}}
}
\def\@without#1{
  \ensuremath{%
    \ifx\hfuzz#1\hfuzz
    \mathrm{d}
    \else
    \foreach \x in {#1}{\mathrm{d}\x\,}
    \fi
    }
}
\makeatother

\title{Line profile of nuclear de-excitation gamma-ray emission from very hot plasma}

\author[H. Yoneda et al.]{
Hiroki Yoneda$^{1,2}$\thanks{E-mail: hiroki.yoneda@uni-wuerzburg.de},
Felix Aharonian$^{3,4,5}$,
Paolo Coppi$^{6}$,
Thomas Siegert$^{1}$, 
Tadayuki Takahashi$^{7,8}$
\\
$^{1}$Julius-Maximilians-Universit\"{a}t W\"{u}rzburg, Fakult\"{a}t f\"{u}r Physik und Astronomie, Institut f\"{u}r Theoretische Physik und Astrophysik, Lehrstuhl f\"{u}r Astronomie, \\ Emil-Fischer-Str. 31, D-97074 W\"{u}rzburg, Germany\\
$^{2}$RIKEN Nishina Center, 2-1 Hirosawa, Wako, Saitama 351-0198, Japan\\
$^{3}$Dublin Institute for Advanced Studies, 31 Fitzwilliam Place, Dublin 2, Ireland\\
$^{4}$Max-Planck-Institut f\"{u}r Kernphysik, P.O. Box 103980, D-69029 Heidelberg, Germany\\
$^{5}$Yerevan State University, 1 Alek Manukyan St, Yerevan 0025, Armenia\\
$^{6}$Yale University, Astronomy Department, P.O. Box 208101 New Haven, CT 06520-8101, USA\\
$^{7}$Kavli Institute for the Physics and Mathematics of the Universe (WPI), University of Tokyo, Kashiwa, Chiba 277-8583, Japan\\
$^{8}$Department of Physics, The University of Tokyo, 7-3-1 Hongo, Bunkyo, Tokyo 113-0033, Japan\\
}

\date{Accepted XXX. Received YYY; in original form ZZZ}

\pubyear{2023}

\begin{document}
\label{firstpage}
\pagerange{\pageref{firstpage}--\pageref{lastpage}}
\maketitle

\begin{abstract}
De-excitation gamma-ray lines, produced by nuclei colliding with protons, provide information about astrophysical environments where particles have kinetic energies of $10-100$ MeV per nucleon.
In general, such environments can be categorized into two types: the interaction between non-thermal MeV cosmic rays and ambient gas, and the other is thermal plasma with a temperature above a few MeV.
In this paper, we focus on the latter type and investigate the production of de-excitation gamma-ray lines in very hot thermal plasma, especially the dependence of the line profile on the plasma temperature.
We have calculated the line profile of prompt gamma rays from $^{12}$C and $^{16}$O and found that when nuclei have a higher temperature than protons, gamma-ray line profiles can have a complex shape unique to each nucleus species. This is caused by anisotropic gamma-ray emission in the nucleus rest frame.
We propose that the spectroscopy of nuclear de-excitation gamma-ray lines may enable to probe energy distribution in very hot astrophysical plasmas. This diagnostics can be a new and powerful technique to investigate the physical state of a two-temperature accretion flows onto a black hole, especially the energy distributions of the protons and nuclei, which are difficult to access for any other diagnostics.
\end{abstract}



\section{Introduction}
\label{sec_intro}
Atomic nuclei colliding with protons can go into their excited states when the kinetic energy of protons is larger than a few MeV in the nucleus rest frame. Immediately after, the excited nuclei emit gamma rays with a specific energy. Such gamma rays, so-called nuclear de-excitation lines, can provide us with unique information about the conditions of production sites of protons and nuclei in the MeV range and their energy distributions and compositions. In astrophysics, understanding the production of such MeV particles is essential in broad topics, for instance, low-energy MeV cosmic rays and their total amount of energy \citep{ramaty_nuclear_1979,Ramaty:1996}, particle acceleration processes in supernova remnants and solar flares \citep{Nobukawa2018,Murphy2009}, and a very hot plasma formed by relativistic compact objects \citep{Aharonian_1984,Ervin:2018a,Ervin:2018b}. Generally, observations of nuclear de-excitation lines can be a direct probe for the following two environments: non-thermal MeV cosmic rays colliding with ambient cold gas and thermal plasma with a temperature above $10^{10}$ K \citep{Aharonian2004}.

The diagnostics of low-energy cosmic rays using the de-excitation lines was noticed with the report of 3--7 MeV gamma-ray excess from the Orion complex observed with COMPTEL/CGRO \citep{Bloemen:1994,Bloemen_1997}. The excess in 3--7 MeV was interpreted as the combination of 4.44 MeV from $^{12}$C$^*$ and 6.13 MeV from $^{16}$O$^*$. 
Furthermore, the reported line was broad, and its width was larger than that of collision between energetic protons in cosmic rays and nuclei in ambient gas (in this case, the line width is $\sim100$ keV). Alternatively, the authors interpreted that non-thermal nuclei in cosmic rays collide with protons in the ambient matter, and the de-excitation lines were Doppler-broadened. While the reported gamma-ray excess was found as a false detection after the background model and analysis method were revisited \citep{Bloemen1999}, it stimulated detailed modeling of the nuclear gamma-ray line due to collisions between non-thermal particles and cold gas \citep{Ramaty_1995,Bykov:1996,Ramaty:1996,Bozhokin_1997}.

The detailed modeling showed that the line profile can have a complex shape under certain situations. \cite{Bykov:1996} and \cite{Kozlovsky:1997} calculated the de-excitation gamma-ray line profiles assuming that nuclei which have kinetic energy larger than $\sim$ 10 MeV nucleon$^{-1}$ collide with ambient protons, and they showed that the line profile could have several peaks. This is caused by the fact that the gamma-ray line emission in the nucleus rest frame is anisotropic, which results in subsequent Doppler broadening of the line shape. In this work, we refer to such a complex line-shape formation as line splitting. The line splitting can also occur when nuclei collide with collimated non-thermal protons, e.g., solar flares \citep{ramaty_nuclear_1979,Murphy_1988,Werntz_1990}. These works revealed that the energy distributions of accelerated protons and nuclei in the MeV range affect the profile of de-excitation lines, and we can derive their energy distributions by the line profile modeling, including the line splitting effect.

While the line profile of de-excitation gamma rays is studied well in the above non-thermal cases, its property under thermal environments still needs to be studied in the same manner. Since the de-excitation gamma rays can also be produced in thermal plasma with a temperature above $10^{10}$ K, e.g., hot plasma surrounding relativistic compact objects and in supernova remnants, understanding the line profile under such thermal environments can provide a unique way to study the two-temperature plasma where electrons and ions have different temperatures and pre-acceleration of low-energy cosmic rays, which cannot be accessible with other tools. Thus, in this work, we aim to examine the basic properties of the line profile of nuclear de-excitation gamma rays under the thermal environments, particularly its dependence on emissivity and proton and nucleus temperatures, and conditions in which the line splitting occurs. 

Here, we mainly focus on the 4.44 MeV and 6.13 MeV gamma-ray lines from $^{12}$C (p, p$\gamma$) $^{12}$C and $^{16}$O (p, p$\gamma$) $^{16}$O. First, we explain the physics of line splitting in Section~\ref{sec_overview} and then describe the method we used to calculate the line profile in Section~\ref{sec_calc}.  Section~\ref{sec_csdata} describes the cross-section data used for the angular distribution of the nuclear gamma rays in the nucleus rest frame.  In Section~\ref{sec_result}, we show the gamma-ray line profile in the thermal environments and describe its basic features without assuming any specific astrophysical situations. In Section~\ref{sec_discussion}, we discuss possible astrophysical applications, such as a unique probe to the two-temperature plasma in accreting flows on black holes. Finally, we summarize our work in Section~\ref{sec_conclusion}.

\section{Line splitting in the profile of prompt gamma-ray line}
\label{sec_overview}
When a nucleus interacts with a proton with enough kinetic energy, the nucleus may be excited into a higher nucleus state. Then, a prompt gamma ray is emitted immediately, with a time scale from fs to ms. The distribution of gamma-ray emission follows multipole radiation corresponding to the spin state of the excited state. Note that here the quantization axis in the nucleus rest frame is determined by the incoming proton direction. In practice, the angular distribution of emitted gamma rays is usually described well with the superposition of several multipole expansion terms because there are several spin states in the excited states \citep{Kolata_1967, ramaty_nuclear_1979}. The angular distribution has been measured for several species experimentally, and it depends on the nuclear species and the projectile energy, that is, here, the kinetic energy of protons \citep{Lesko:1988,Lang:1987}.

\begin{figure}
\begin{center}
\includegraphics[width=8 cm]{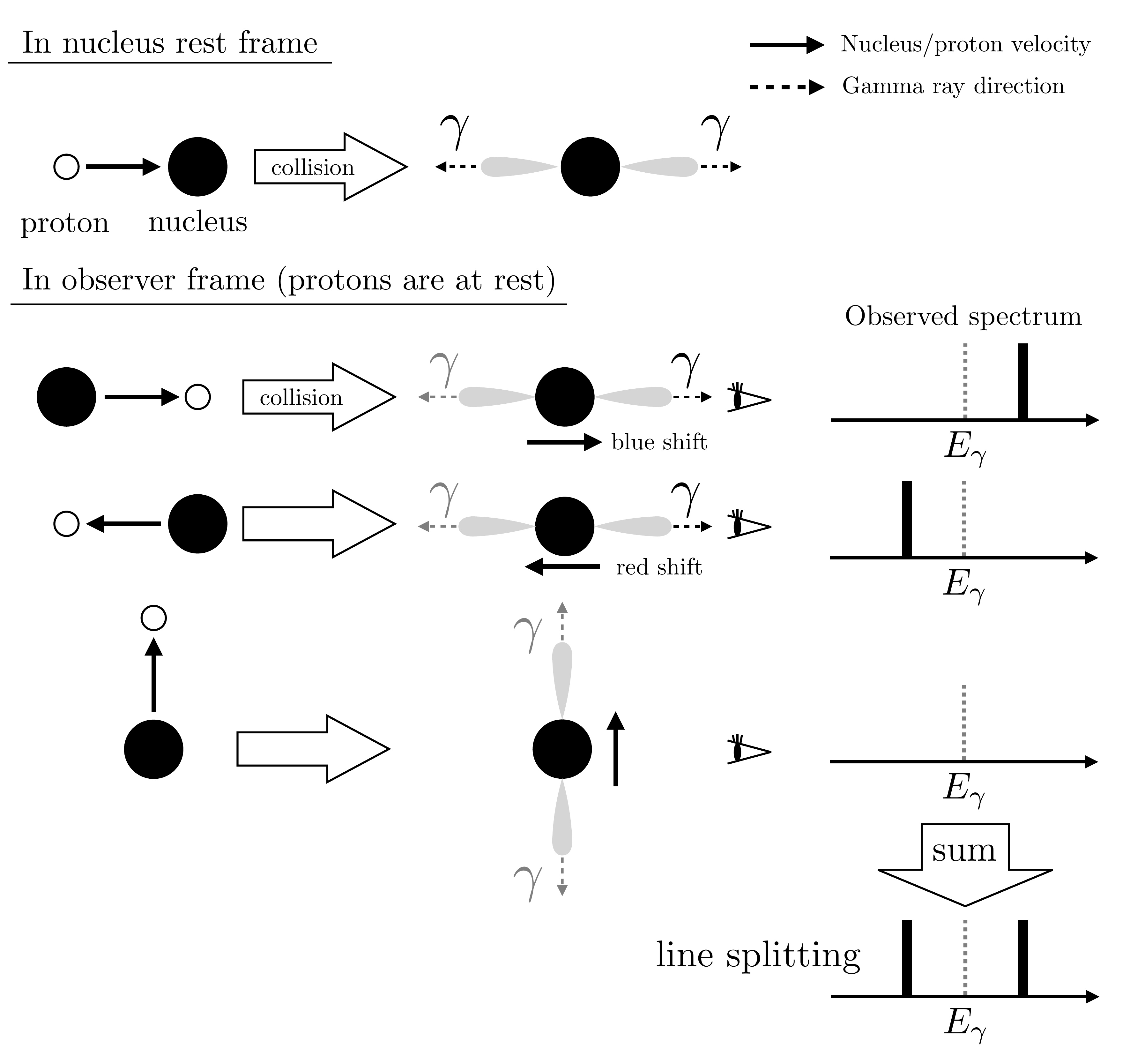}
\caption{A schematic of line splitting. For simplicity, here we assume that the nuclei emit gamma rays in only two directions in the nucleus rest frame (top). In the observer frame (bottom), the observer can detect gamma rays from only the top two cases, which results in a split line profile.}
\label{fig_line_splitting_overview}
\end{center}
\end{figure}

Suppose a nucleus has a kinetic energy larger than an excited state, so that it could change the states when interacting with a proton at rest. For this self-excitation, the nucleus's energy has to be several tens of MeV per nucleon so that the proton's energy can be larger than the nucleus's excited state at the nucleus rest frame. Then, the nucleus's velocity is several tens percent of the speed of light. Therefore, the prompt gamma ray is emitted from the fast-moving nucleus, and its energy is Doppler-shifted, corresponding to the nucleus's velocity relative to the observer. Considering that the quantization axis of the gamma-ray emission is the incoming proton direction in the nucleus rest frame, it is aligned approximately with the nucleus velocity vector when converted into the observer frame. Figure~\ref{fig_line_splitting_overview} visualizes this situation. Here we assume that the nucleus emits gamma rays in only two directions, as the simplest case of the anisotropic emission in the nucleus rest frame. Due to the anisotropy, the gamma-ray emission can be aligned with the nucleus velocity in the observer frame. In the figure, the gamma rays can be observed only when the nuclei move toward or away from the observer, resulting in the split line profile with two peaks. Generally, when the gamma-ray emission is anisotropic in the nucleus rest frame, gamma rays with a certain Doppler shift can be suppressed. Then, if the nucleus has large kinetic energy, the observed gamma-ray line shape can be complex, e.g., being split, reflecting its original anisotropic emission. 

The actual profile depends on the energy distributions of particles, the collisional cross section, and the angular distribution of anisotropic gamma-ray emission at each energy. In the following section, we explain the procedure of calculating the prompt gamma-ray line shape with cross-section data and arbitrary particle energy distributions.

\section{Gamma-ray line profile calculation}
\label{sec_calc}
The reaction rate ($R$) per unit volume between two different particle species (here, we consider protons and nuclei) is described as \citep{Weaver:1976}
\begin{eqnarray}
R = c n_{\mathrm{A}} n_{\mathrm{p}} \iint \dif{\bm{p}_{\mathrm{A}},\bm{p}_{\mathrm{p}}} f(\bm{p}_{\mathrm{A}}) f(\bm{p}_{\mathrm{p}}) \frac{\gamma'_{\mathrm{p}}\beta'_{\mathrm{p}}}{\gamma_{\mathrm{p}}\gamma_{\mathrm{A}}}\sigma(\gamma'_{\mathrm{p}})~,
\end{eqnarray}
where $c$ is the speed of light, $n_{\mathrm{A}/\mathrm{p}}$ are the number densities of nuclei and protons in the observer frame, $\bm{p}$ are the momentum vectors; $f(\bm{p}_{\mathrm{A/\mathrm{p}}})$ are the momentum distributions of nuclei and protons in the observer frame; $\beta$ and $\gamma$ are the ratio of velocity to $c$ and the Lorentz factor, respectively. The primed variables indicate those in the nucleus rest frame, and the non-primed variables indicate those in the observer frame. $\sigma(\gamma'_{\mathrm{p}})$ is the collisional cross section for the excitation process. To calculate the line profile observed from a certain direction, we need to implement the anisotropy of the gamma-ray emission in the nucleus rest frame. To do so, we expand $\sigma(\gamma'_{\mathrm{p}})$ as 
$
\displaystyle \int \dif{\Omega_{\gamma}} \dif[\Omega'_{\gamma}]{\Omega_{\gamma}}   \dif[\sigma\left(\gamma'_{\mathrm{p}}{, \ }\theta'_{\gamma \mathrm{p}}\right)]{\Omega'_{\gamma}}
$.
Note that $\Omega_{\gamma}$ and $\Omega'_{\gamma}$ are the solid angles in the observer and nucleus rest frames, respectively, and $\theta'_{\gamma \mathrm{p}}$ is the angle between the gamma-ray and incoming proton directions in the nucleus rest frame. In Section~\ref{sec_csdata}, we will outline our database used for 
$
\displaystyle \dif[\sigma\left(\gamma'_{\mathrm{p}}{, \ }\theta'_{\gamma \mathrm{p}}\right)]{\Omega'_{\gamma}}
$.
As shown in \cite{RadiPro}, we use
\begin{eqnarray}
\dif[\Omega'_{\gamma}]{\Omega_{\gamma}} = \frac{1}{\gamma_{\mathrm{A}}^2(1 - \beta_{\mathrm{A}} \cos\theta_{\gamma \mathrm{A}})^2}~,
\end{eqnarray}
where $\theta_{\gamma \mathrm{A}}$ is the angle between the directions of the gamma ray and the incoming nucleus in the observer frame. Then, we can derive the photon emissivity in the observer's direction
\begin{eqnarray}
\label{eq_core}
\begin{aligned}
\frac{\dif{R}}{\dif{\Omega_{\gamma}}} = c n_{\mathrm{A}} & n_{\mathrm{p}} \iint \dif{\bm{p}_{\mathrm{A}},\bm{p}_{\mathrm{p}}}  f(\bm{p}_{\mathrm{A}}) f(\bm{p}_{\mathrm{p}}) \\
& \times \frac{\gamma'_{\mathrm{p}}\beta'_{\mathrm{p}}}{\gamma_{\mathrm{p}}\gamma_{\mathrm{A}}}\frac{1}{\gamma_{\mathrm{A}}^2(1 - \beta_{\mathrm{A}} \cos\theta_{\gamma \mathrm{A}})^2}\dif[\sigma\left(\gamma'_{\mathrm{p}}{, \ }\theta'_{\gamma \mathrm{p}}\right)]{\Omega'_{\gamma}}~.
\end{aligned}
\end{eqnarray}

To compute the integral in Equation~\ref{eq_core}, we adopt Monte Carlo integration by sampling $\bm{p}_{\mathrm{p}}$ and $\bm{p}_{\mathrm{A}}$, and calculate the gamma-ray line profile numerically. The following is the procedure for the calculation.
\begin{enumerate}
\item We sample the proton momentum, $\bm{p}_{\mathrm{p}}$, in the observer frame and calculate its energy $E_{\mathrm{p}}$.
The momentum distribution is assumed to be isotropic. In the case of thermalized protons, we sample the momentum component first and then calculate the energy so that the momentum components obey Maxwellian distribution.
\begin{eqnarray}
& f(p_{\mathrm{p},i}) \propto \displaystyle \exp\left(-\frac{p_{\mathrm{p},i}^2}{2 m_{\mathrm{p}} k T_{\mathrm{p}}}\right)~,i=1,2,3\\
& E_{\mathrm{p}} = \sqrt{\left(\displaystyle\sum_{i=1,2,3}c^2 p_{\mathrm{p},i}^2\right) + m_\mathrm{p}^2 c^4}~.
\end{eqnarray}
Note that $m_{\mathrm{p}}$ and $T_{\mathrm{p}}$ are the proton mass and temperature, respectively; $k$ is the Boltzmann constant.
\item In the same way as step (i), we sample the nucleus momentum, $\bm{p}_{\mathrm{A}}$ and calculate its energy $E_{\mathrm{A}}$.
Its momentum distribution can be defined as
\begin{eqnarray}
& f(p_{\mathrm{A},i}) \propto \displaystyle\exp\left(-\frac{p_{\mathrm{A}, i}^2}{2 m_{\mathrm{A}} k T_{\mathrm{A}}}\right)~,i=1,2,3\\
& E_{\mathrm{A}} = \sqrt{\left(\displaystyle\sum_{i=1,2,3}c^2 p_{\mathrm{A},i}^2\right) + m_{\mathrm{A}}^2 c^4}~,
\end{eqnarray}
where $m_{\mathrm{A}} (= A \times m_p)$ and $T_{\mathrm{A}}$ are the nucleus mass and temperature, respectively. Note that $A$ is the mass number of the nucleus.
\item We fix the gamma-ray direction as:
\begin{eqnarray}
\bm{p}_{\gamma} \propto (0,0,1)
\end{eqnarray}
Note that since the particle momentum distributions are isotropic here, the resulting gamma rays are also isotropic. Thus, the result does not depend on the observer's direction. The (Doppler-shifted) gamma-ray energy $E_{\gamma}$ will be determined at step (v).
\item We convert $(E_{\mathrm{p}}, \bm{p}_{\mathrm{p}})$ and $(E_{\mathrm{\gamma}}, \bm{p}_{\gamma})$ into 4-momenta in the nucleus rest frame, $(E'_{\mathrm{p}}, \bm{p}'_{\mathrm{p}})$ and $(E'_{\gamma}, \bm{p}'_{\gamma})$ via Lorentz transformation.
Then, we calculate a weight $w$, which we use to fill the gamma-ray energy of each event into a histogram.
\begin{eqnarray}
& w = \displaystyle \frac{\gamma'_{\mathrm{p}}\beta'_{\mathrm{p}}}{\gamma_{\mathrm{p}}\gamma_{\mathrm{A}}}\frac{1}{\gamma_{\mathrm{A}}^2(1 - \beta_{\mathrm{A}} \cos\theta_{\gamma \mathrm{A}})^2}\dif[\sigma\left(\gamma'_{\mathrm{p}}{, \ }\theta'_{\gamma \mathrm{p}}\right)]{\Omega'_{\gamma}}\\
& \displaystyle \gamma'_{\mathrm{p}} = \frac{p^{\mu}_{\mathrm{p}}p_{\mathrm{A}\mu}}{m_\mathrm{A} m_\mathrm{p} c^2}
\end{eqnarray}
\item We consider the gamma-ray energy in the nucleus rest frame, $E'_{\gamma}$, to be fixed.
For example, the energy of gamma rays from $^{12}$C is 4.44 MeV. Then, we again convert $(E'_{\gamma}, \bm{p}'_{\gamma})$ into $(E_{\gamma}, \bm{p}_{\gamma})$ in order to calculate the observed gamma-ray energy, $E_{\gamma}$.
\begin{eqnarray}
\left(E'_{\gamma}, c p'_{1}, c p'_{2}, c p'_{3}\right) \longmapsto 
\left(E_{\gamma},0,0,E_{\gamma}\right)
\end{eqnarray}
In this step, we ignore the energy shift due to the nucleus recoil after the proton collision. As discussed in Section 2 of \cite{Kozlovsky:1997}, this recoil energy can affect the line shape if the nuclei are at rest and the protons have large energy. However, in this work, the nuclei have large velocities, and the Doppler shift by their motions is dominant. Thus, the recoil effect is insignificant under the conditions of this work. In Appendix~\ref{sec_recoil}, we calculate the gamma-ray spectra considering the recoiling energy and evaluate this approximation. While the peak height in the spectra can be modified by a few \% under special conditions, the basic features of the line profile are not affected by the recoil effect.
\item Finally, we fill $E_{\gamma}$ calculated in step (v) into a histogram with the weight $w$.
\item We repeat the steps (i)-(iv) until sufficient statistics are accumulated and normalize the histogram by the number of Monte Carlo samples. Typically, $10^7$ samples are used for each line profile calculation in this work.
\end{enumerate}

In Section~\ref{sec_result}, following the above procedure, we will calculate the line gamma-ray emissivity per unit volume and density, defined as 
\begin{eqnarray}
Q_{\gamma} = \frac{1}{n_{\mathrm{A}} n_{\mathrm{p}}} \frac{\dif{R}}{\dif{\Omega_{\gamma}}}~.
\end{eqnarray}
We will also show the gamma-ray line profile, which can be derived as 
\begin{eqnarray}
\frac{\dif{Q_{\gamma}}}{\dif{E_{\gamma}}}~.
\end{eqnarray}

\section{Cross section data}
\label{sec_csdata}
For our calculations, we describe our dataset for the angular distribution of the nuclear gamma-ray emission in the excited nucleus rest frame. The gamma-ray direction $\mathrm{d}\Omega'_{\gamma}$ is determined relative to the momentum direction of an initial incoming proton. The differential cross section $\displaystyle \dif[\sigma]{\Omega'_{\gamma}}$ is usually described well with Legendre polynomials
\begin{eqnarray}
\label{eq_def_cs}
\begin{split}
\dif[\sigma (E'_{\mathrm{K,p}}{, \ }\theta')]{\Omega'_{\gamma}} = \frac{\sigma(E'_{\mathrm{K,p}})}{4\upi}\sum\limits_{l=0}^{l_\textsc{max}}b_{2l}(E'_{\mathrm{K,p}}) P_{2l}(\cos\theta')~,\\
\mathrm{where}~b_0 = 1~.
\end{split}
\end{eqnarray} 
Note that here the cross section is parameterized with the kinetic energy of protons in the nucleus rest frame $E'_{\mathrm{K,p}} (= E'_{\mathrm{p}} - m_{\mathrm{p}}c^2)$.  

\begin{figure}
    \begin{center}
    \includegraphics[width=8.0 cm]{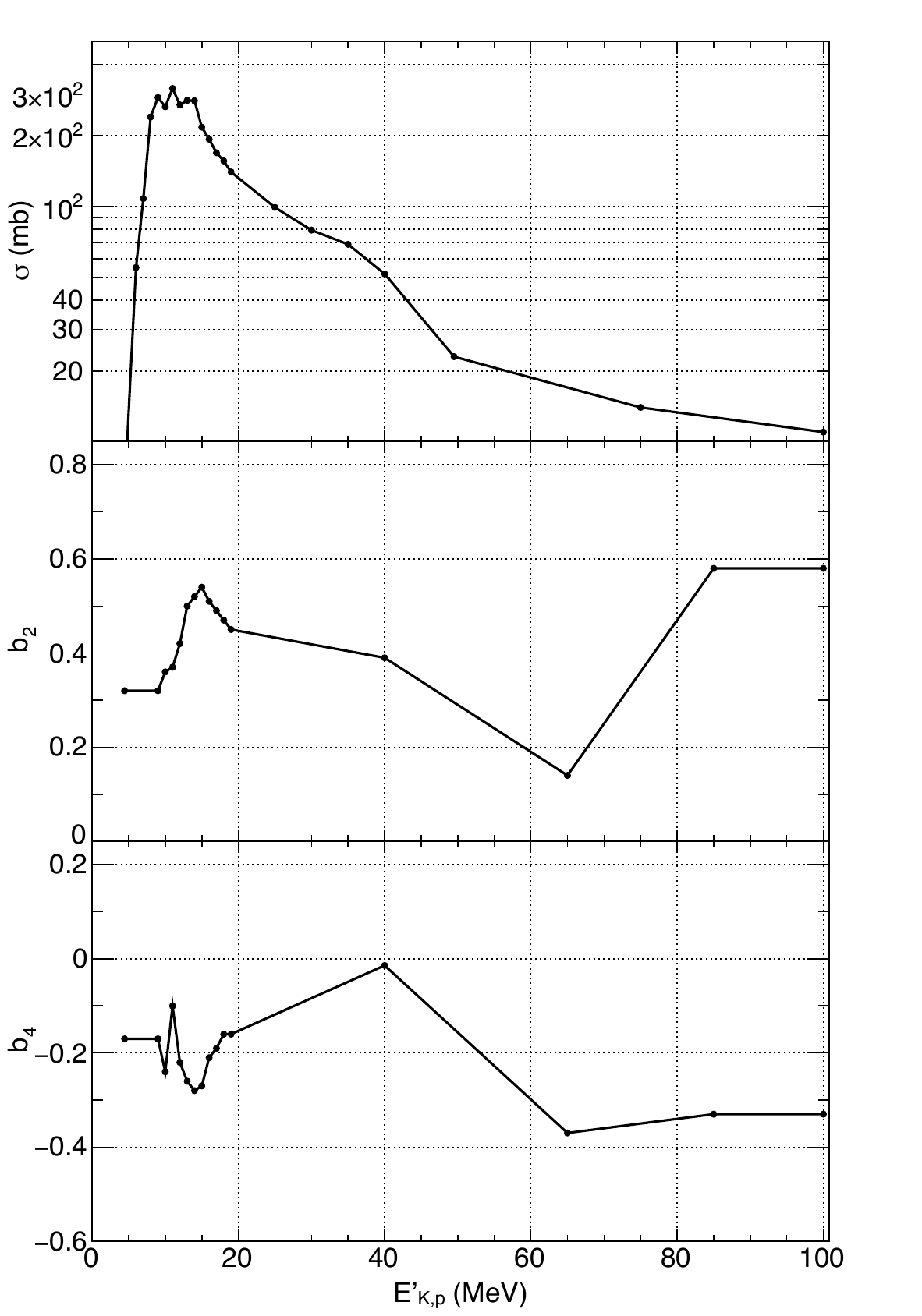}
    \includegraphics[width=8.0 cm]{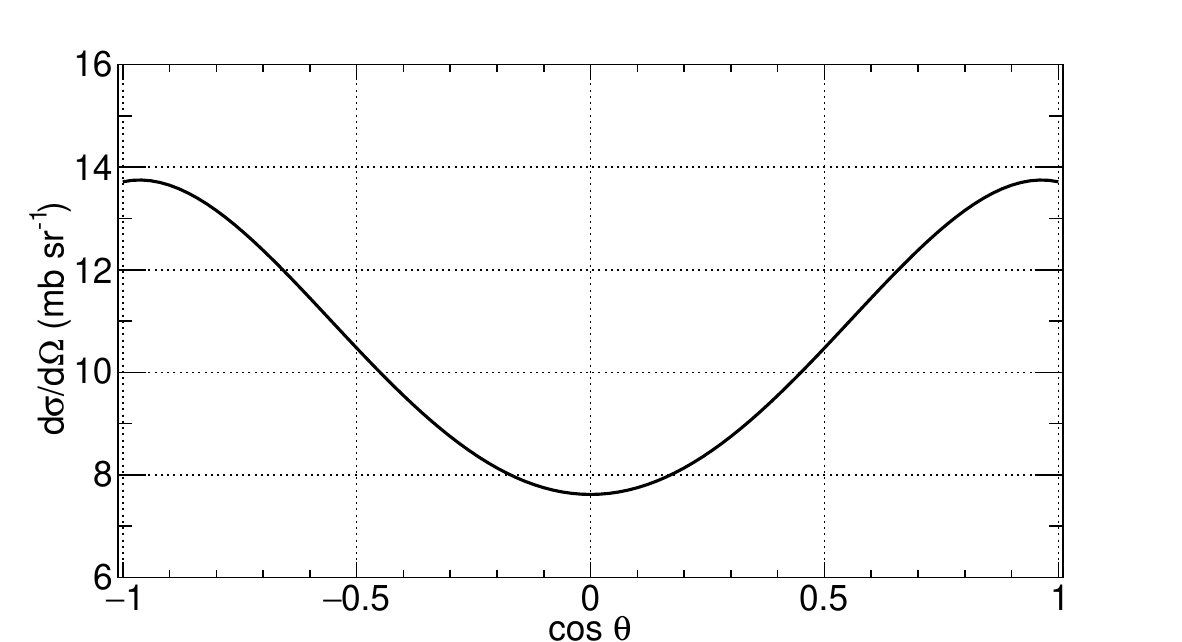}
    \caption{The cross section data which we use for 4.44 MeV gamma-ray emission from $^{12}$C (p, p$\gamma$) $^{12}$C. The top is the total cross section. The second and third are the coefficient $b_2$ and $b_4$, respectively. The bottom is the differential cross section at $E'_{\mathrm{K,p}} = 20$ MeV as an example.}
    \label{fig:cs_carbon}
    \end{center}
\end{figure}

We extracted the cross-section data from several experimental works in nuclear physics (for $^{12}$C (p, p$\gamma$) $^{12}$C, \citet{Kiener:1998,Lang:1987,Kiener:2001,Lesko:1988}, for $^{16}$O (p, p$\gamma$) $^{16}$O, \citet{Kiener:1998,Lang:1987,Lesko:1988}). Figure~\ref{fig:cs_carbon} shows the data we use for the 4.44 MeV gamma ray line from $^{12}$C interactions with protons. Appendix figure~\ref{fig:cs_oxygen} shows those for 6.13 MeV gamma rays from $^{16}$O interactions with protons. Using the extracted data points, we calculate the functions $b_{2l}$ and the total cross section with linear interpolation.

For $^{12}$C, due to the lack of experimental data, we assume that $\sigma(E'_{\mathrm{K,p}}=4.44~\mathrm{MeV}) = 0$ mb, and the functions $b_2, b_4$ between 4.44 and 9 MeV are the same as those measured at 9 MeV. We check the validity of this approximation in Appendix~\ref{sec:approximation_validity}. The cross section at $>$ 100 MeV is calculated as $\displaystyle \left(E'_{\mathrm{K,p}}/100~\mathrm{MeV}\right)^{-1} \times \sigma(E'_{\mathrm{K,p}}=100~\mathrm{MeV})$, and the coefficients are the same as those at 100 MeV. For $^{16}$O, also due to the lack of experimental data, we assume that $\sigma(E'_{\mathrm{K,p}}=6.13~\mathrm{MeV}) = 0$ mb, and the functions $b_2, b_4, b_6$ between 6.13 and 9 MeV are the same as those measured at 9 MeV. The cross section at $>$ 85 MeV is calculated as $\displaystyle \left(E'_{\mathrm{K,p}}/85~\mathrm{MeV}\right)^{-1} \times \sigma(E'_{\mathrm{K,p}}=85~\mathrm{MeV})$, and the coefficients are the same as those at 85 MeV.

\begin{figure}
    \begin{center}
    \includegraphics[width=8.5 cm]{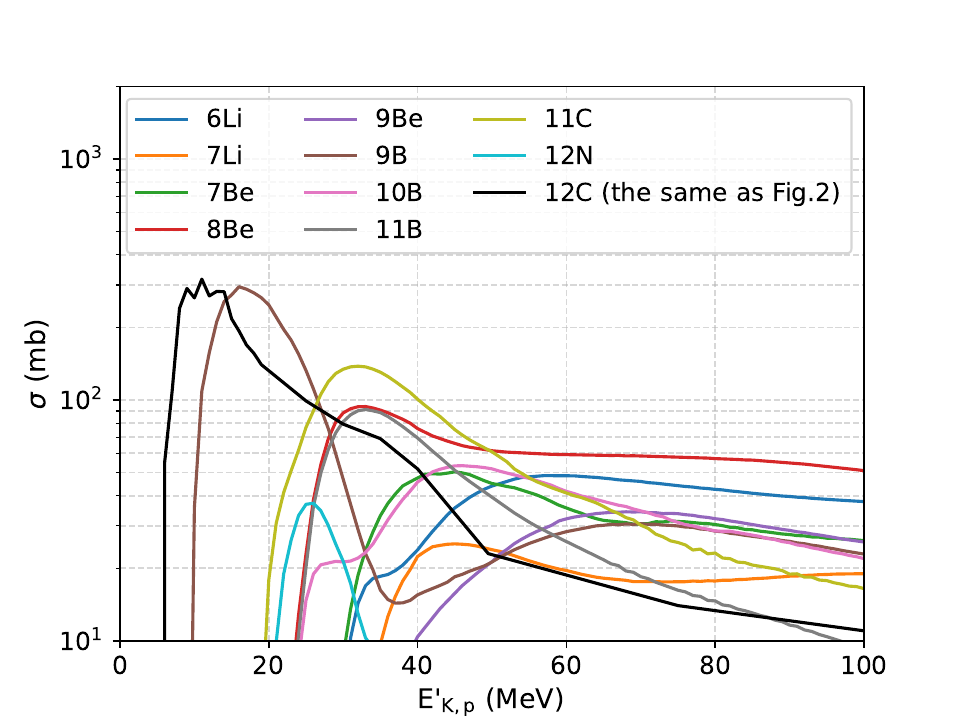}
    \caption{The cross sections of the spallation processes of Carbon based on the TALYS code. The legend displays produced nuclei in the spallation processes. Here only the processes with the maximum cross section larger than 20 mb are shown.}
    \label{fig:cs_comp}
    \end{center}
\end{figure}

It should be noted that the spallation process starts to dominate over the excitation above $\sim$ 20 MeV.
We ran the nuclear reaction simulator TALYS \citep{TALYS2005} and compared spallation cross sections of Carbon with its de-excitation process. Figure~\ref{fig:cs_comp} shows that the cross sections of the spallation processes producing $^{9}$B, $^{11}$C, and $^{8}$Be can be larger than the de-excitation process in a certain energy range. The spallation process could reduce the amount of Carbon and its line emissivity. On the other hand, the spallation of heavier nuclei would produce Carbon. For example, $^{16}$O (p, x $\gamma_{4.4~\mathrm{MeV}}$) $^{12}$C has a comparable cross section of $^{16}$O (p, $\gamma$) $^{16}$O. This process can produce a 4.44 MeV gamma ray and increase the amount of Carbon after that. While it is beyond the scope of this paper, such a nuclear reaction network should be considered in the line modeling of certain astronomical sources.

\section{Result}
\label{sec_result}
To focus on the basic properties of the gamma-ray line shape,
we investigate it by changing the proton and nucleus temperatures, which are represented as $kT_{\mathrm{p}}$ and $kT_{\mathrm{A}}$, respectively. Then, we will discuss its astrophysical applications in the following section.
Here we mainly show the line profile of 4.44 MeV due to $^{12}$C (p, p$\gamma$) $^{12}$C.
Since basic features are found to be the same as those of $^{12}$C,
the results for the 6.13 MeV gamma-ray line from $^{16}$O are shown in Appendix~\ref{sec:oxygen}.

\begin{figure}
\begin{center}
\includegraphics[width=8.0 cm]{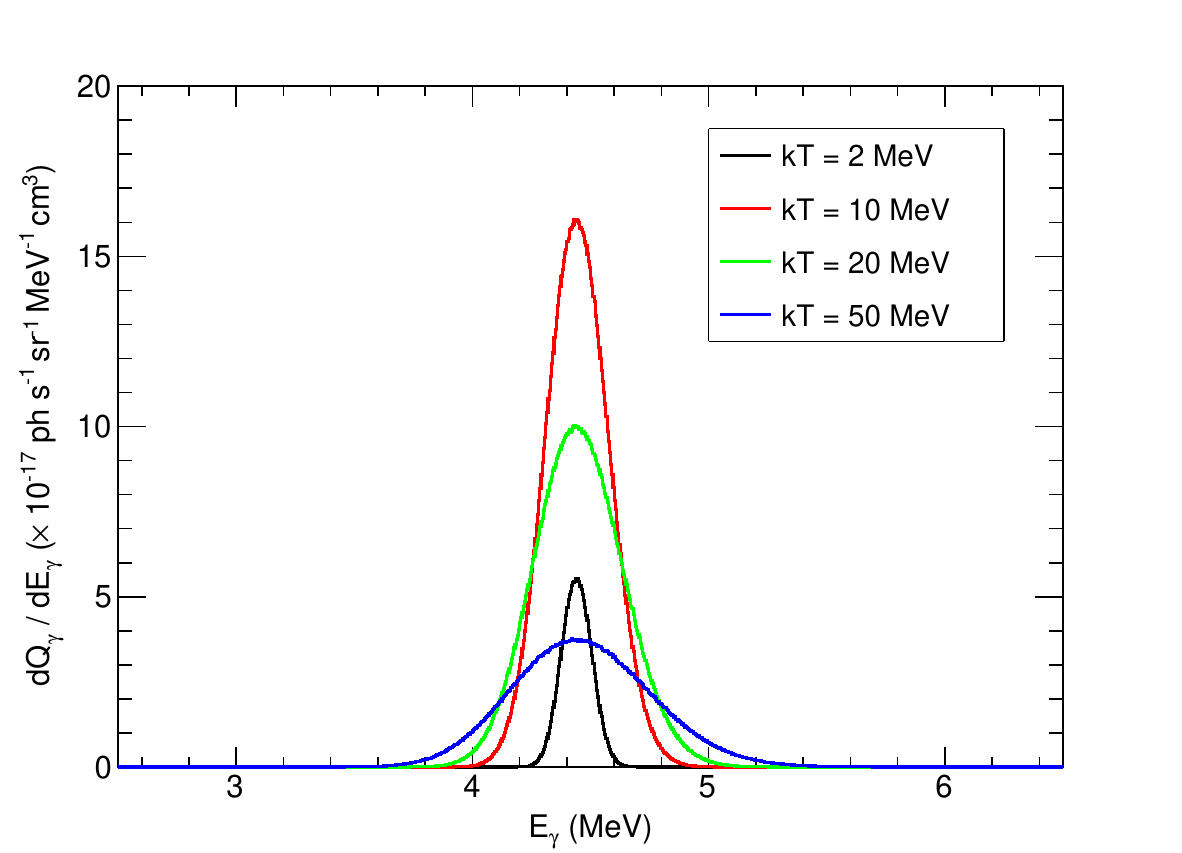}
\caption{4.44 MeV gamma-ray line shape in the case that $T_{\mathrm{p}} = T_{\mathrm{A}}$. Black: $kT_{\mathrm{A}} = kT_{\mathrm{p}} = 2.0~\mathrm{MeV}$, Red: $10.0~\mathrm{MeV}$, Green: $20.0~\mathrm{MeV}$, Blue: $50.0~\mathrm{MeV}$.}
\label{fig:lineshape_sameTemp_carbon}
\end{center}
\end{figure}

\begin{figure}
    \begin{center}
    \includegraphics[width=8.0 cm]{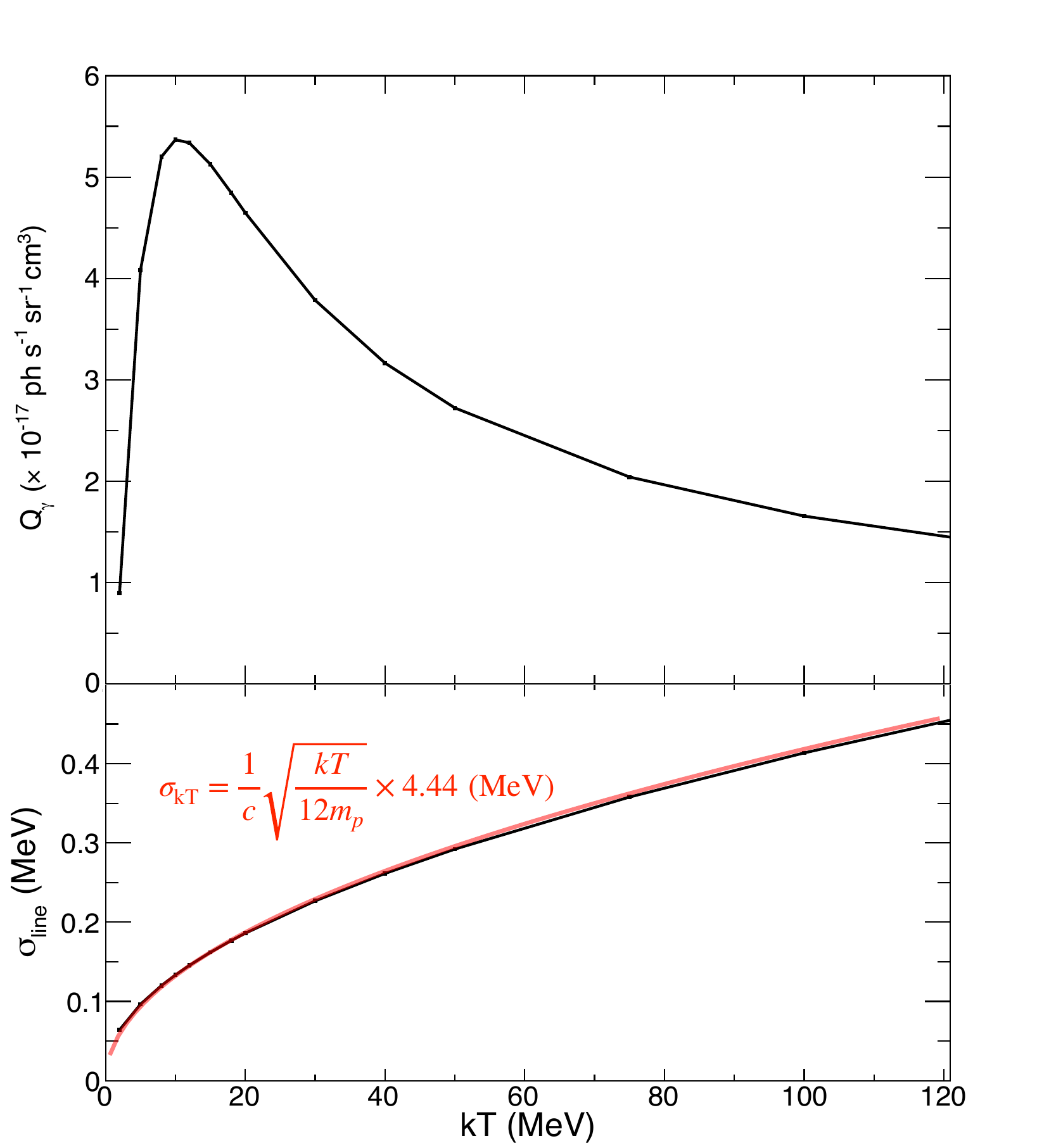}
                    \caption{The dependence of the emissivity and the line width on the temperature in the case that $T_\mathrm{p} = T_\mathrm{A}$. The top panel shows the gamma-ray emissivity depending on the temperature, and the bottom is the standard deviation of the line profile. The red line is the value expected from the normal thermal Doppler broadening.}
    \label{fig:photonflux_sameTemp_carbon}
        \end{center}
\end{figure}

First, we calculate the line profile under equal temperatures, $T_{\mathrm{p}}$ and $T_{\mathrm{A}}$.
We vary the common temperature from 2 MeV to 50 MeV. Figure~\ref{fig:lineshape_sameTemp_carbon} shows the calculated line profiles. In this case, the line profile is described well by a Gaussian, and the line splitting does not occur significantly. Figure~\ref{fig:photonflux_sameTemp_carbon} shows the photon emissivity and the standard deviation $\sigma_{\mathrm{line}}$ of the line profile at each temperature. By tracing the cross section of $^{12}$C (p, p$\gamma$) $^{12}$C, the emissivity obtains a maximum around 10 MeV. Also, the line width is consistent with the normal thermal Doppler broadening of
\begin{eqnarray}
\label{eq_sigma_thermal}
\sigma_{kT_{\mathrm{A}}} = \sqrt{\frac{kT_{\mathrm{A}}}{m_{\mathrm{A}} c^2}} E_{\gamma}~.
\end{eqnarray}

\begin{figure}
    \begin{center}
    \includegraphics[width=8.0 cm]{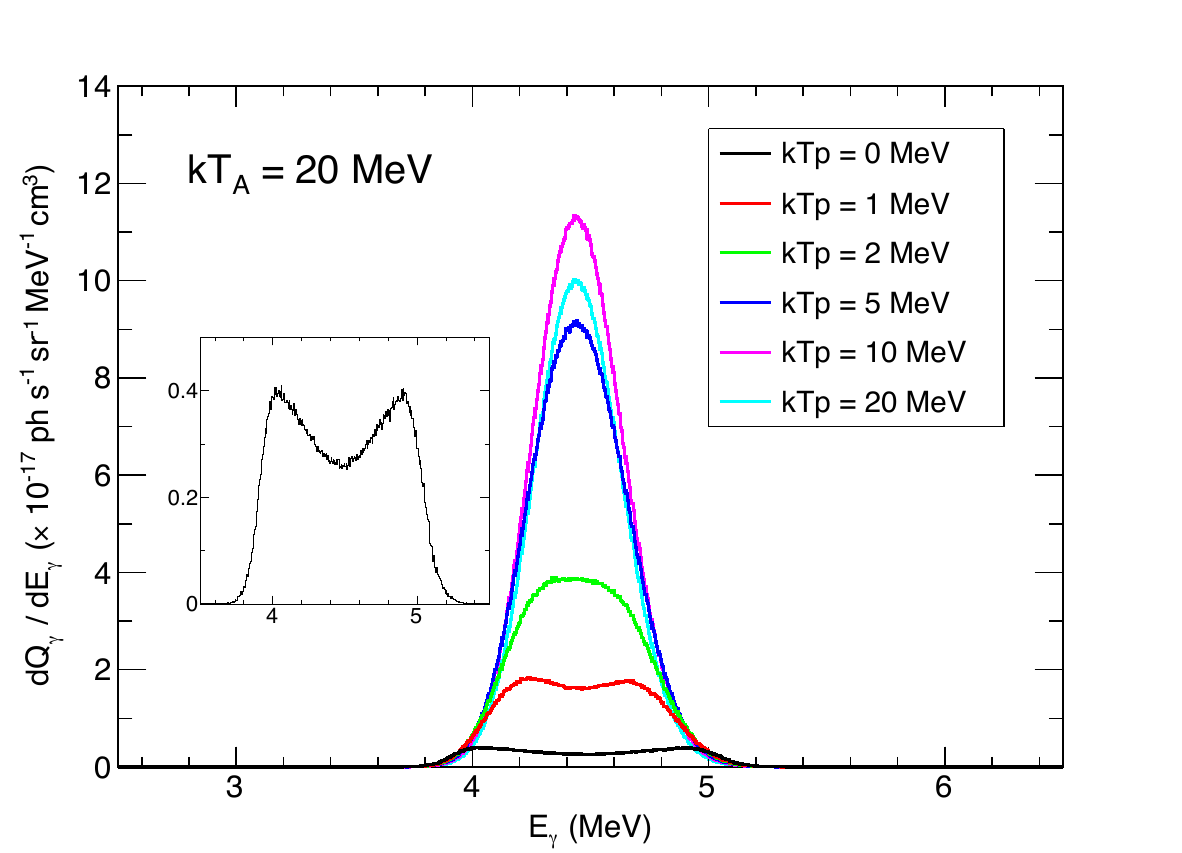}
    \includegraphics[width=8.0 cm]{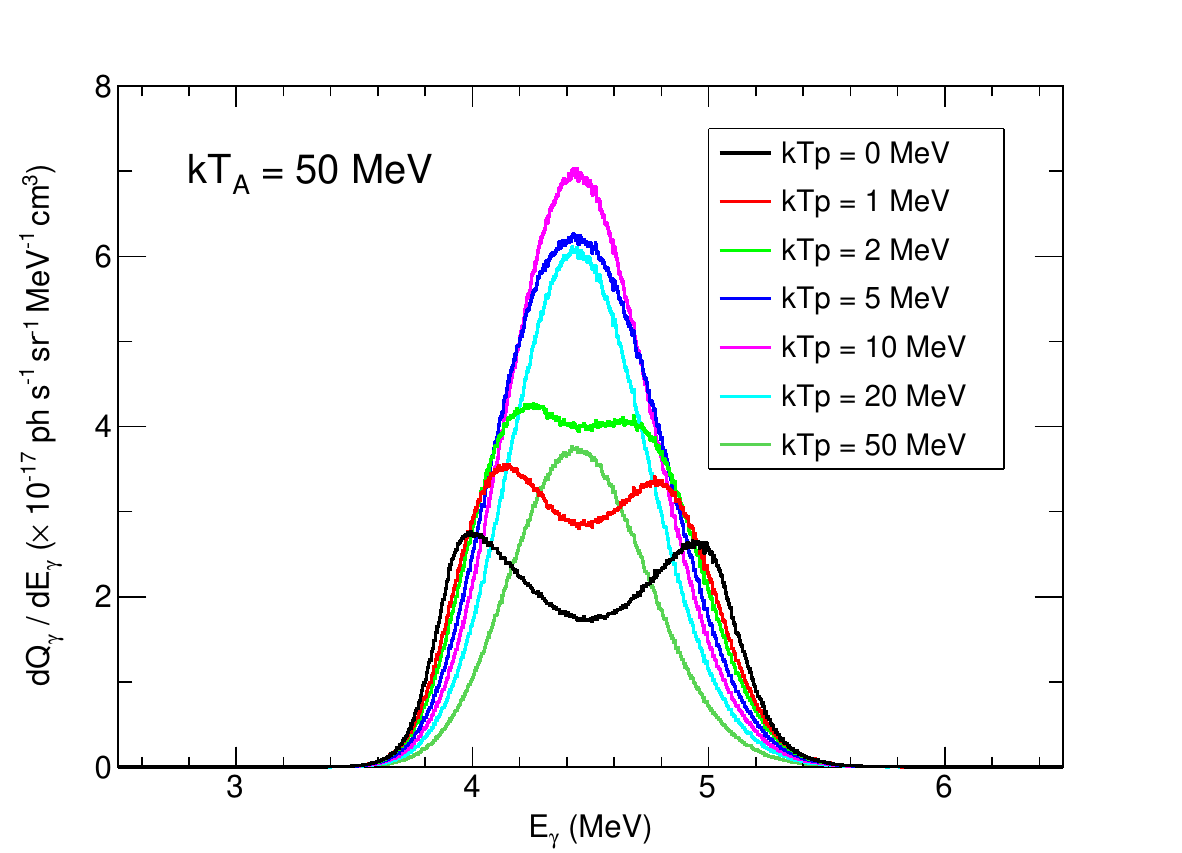}
    \caption{The line profile of 4.44 MeV gamma-ray lines with different proton temperatures. The nucleus temperature is fixed to 20 MeV (top) or 50 MeV (bottom). The line splitting is clearly seen when $kT_\mathrm{p}$ is less than a few MeV. The inset in the top panel is a zoom of the line profile with $kT_\mathrm{p}$ = 0 MeV.}
    \label{fig:lineshape_kTA_carbon}
    \end{center}
\end{figure}

Next, as a typical case where the line splitting occurs, we show the line profile when $T_{\mathrm{A}}$ is significantly higher than $T_{\mathrm{p}}$. Figure~\ref{fig:lineshape_kTA_carbon} shows the line profile by fixing $kT_{\mathrm{A}}$ to 20 or 50 MeV, and changing $kT_{\mathrm{p}}$ from 0 MeV to $kT_{\mathrm{A}}$. As seen clearly, when $kT_{\mathrm{p}}$ is lower than a few MeV, the line profile is distorted from a Gaussian shape, and it has two peaks by reflecting the anisotropic gamma-ray emission in the nucleus rest frame as seen in Figure~\ref{fig:cs_carbon}. As $kT_{\mathrm{p}}$ becomes larger than $\sim$ 5 MeV, the line profile is closer to a Gaussian-like shape again. This is because as far as the proton's kinetic energy gets close to or less than the nucleus's kinetic energy per nucleon, i.e., $kT_{\mathrm{p}} \sim kT_{\mathrm{A}} / A$, the quantization axis of the gamma-ray emission is nearly parallel to the nucleus momentum direction in the observer frame, which causes the line splitting. However, when the proton's kinetic energy becomes larger, the quantization axis is less aligned with the nucleus momentum direction and randomized in the observer frame. Then, the anisotropy is smeared out, and the line becomes Gaussian again. Considering that the mass number of carbon is 12, the maximum proton temperature which yields the line splitting can be estimated as $\sim$2 (=20/12) to $\sim$4 (=50/12) MeV, which is consistent with Figure~\ref{fig:lineshape_kTA_carbon}. In Appendix~\ref{sec:oxygen}, we also show the line profile of 6.13 MeV from $^{16}$O. Since the mass number of oxygen is 16, larger than carbon, the line splitting is smeared out at a lower proton temperature than for carbon (see Figure~\ref{fig:lineshape_kTA_oxygen}).

To make the dependence of the line profile on the hot plasma temperatures clear, we performed 
a parameter scan for $T_{\mathrm{p}}$ and $T_{\mathrm{A}}$. For each, we assigned values from 0.5 MeV to 100 MeV. In addition to the gamma-ray emissivity $Q_{\gamma}$ and the line width $\sigma_{\mathrm{line}}$, we also investigate the line width normalized by the normal thermal Doppler broadening width $\sigma_{kT_{\mathrm{A}}}$ because it is a good indicator to know how much the line profile deviates from a Gaussian shape determined by the nucleus temperature. As shown in Figure~\ref{fig:photonflux_sameTemp_carbon}, $\sigma_{\mathrm{line}} / \sigma_{kT_{\mathrm{A}}}$ is equal to one when the line splitting is not significant.

\begin{figure}
    \begin{center}
    \includegraphics[width=8.0 cm]{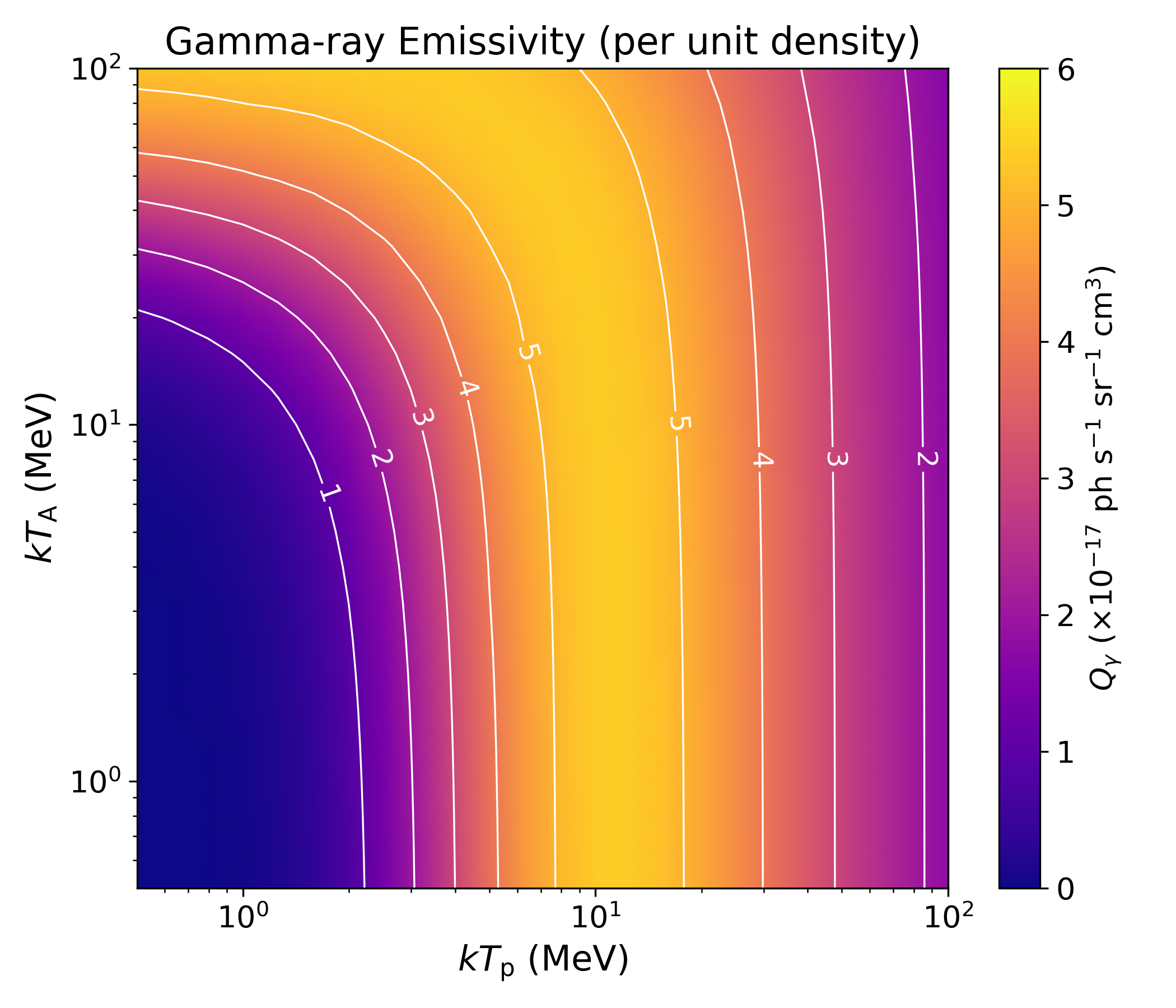}
    \includegraphics[width=8.0 cm]{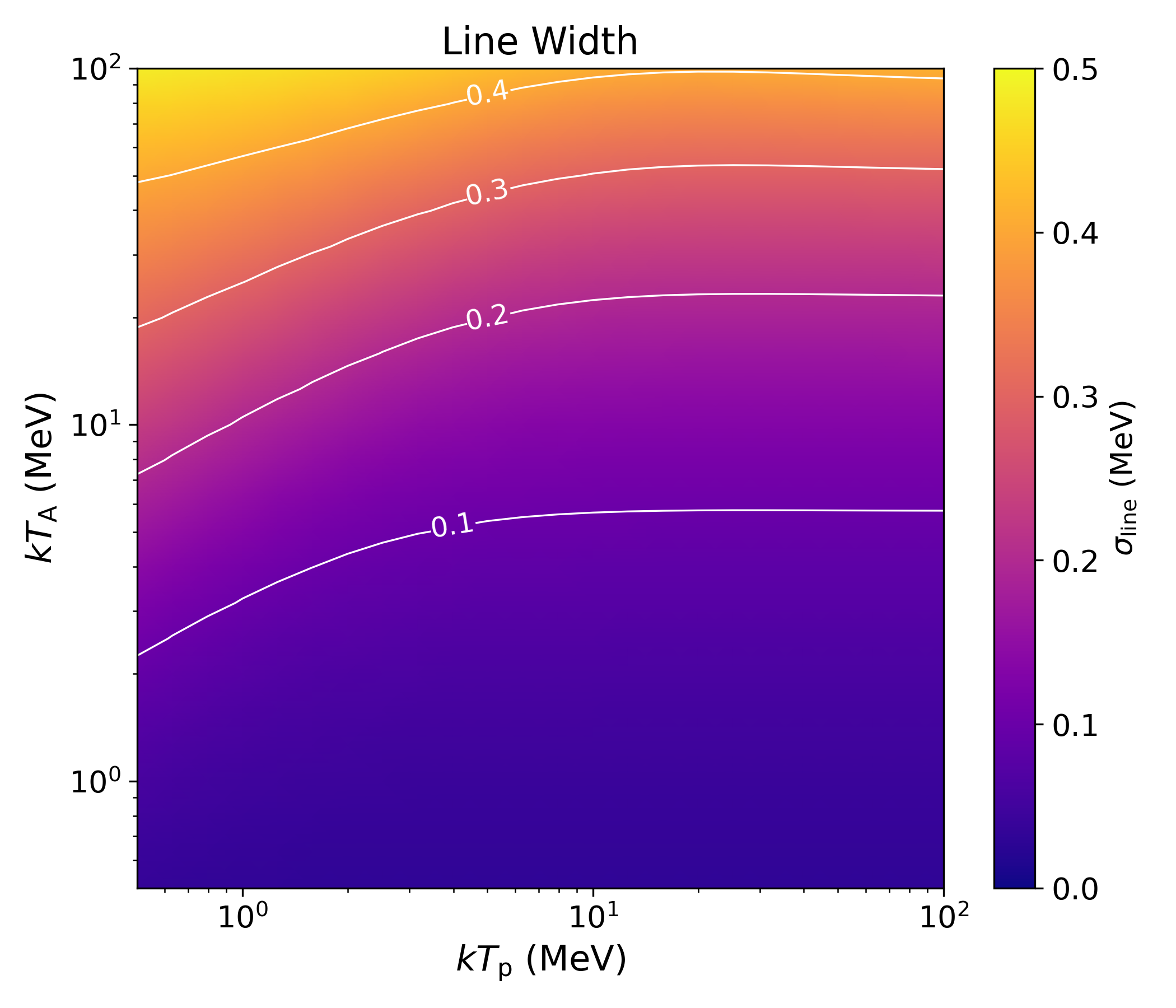}
    \includegraphics[width=8.0 cm]{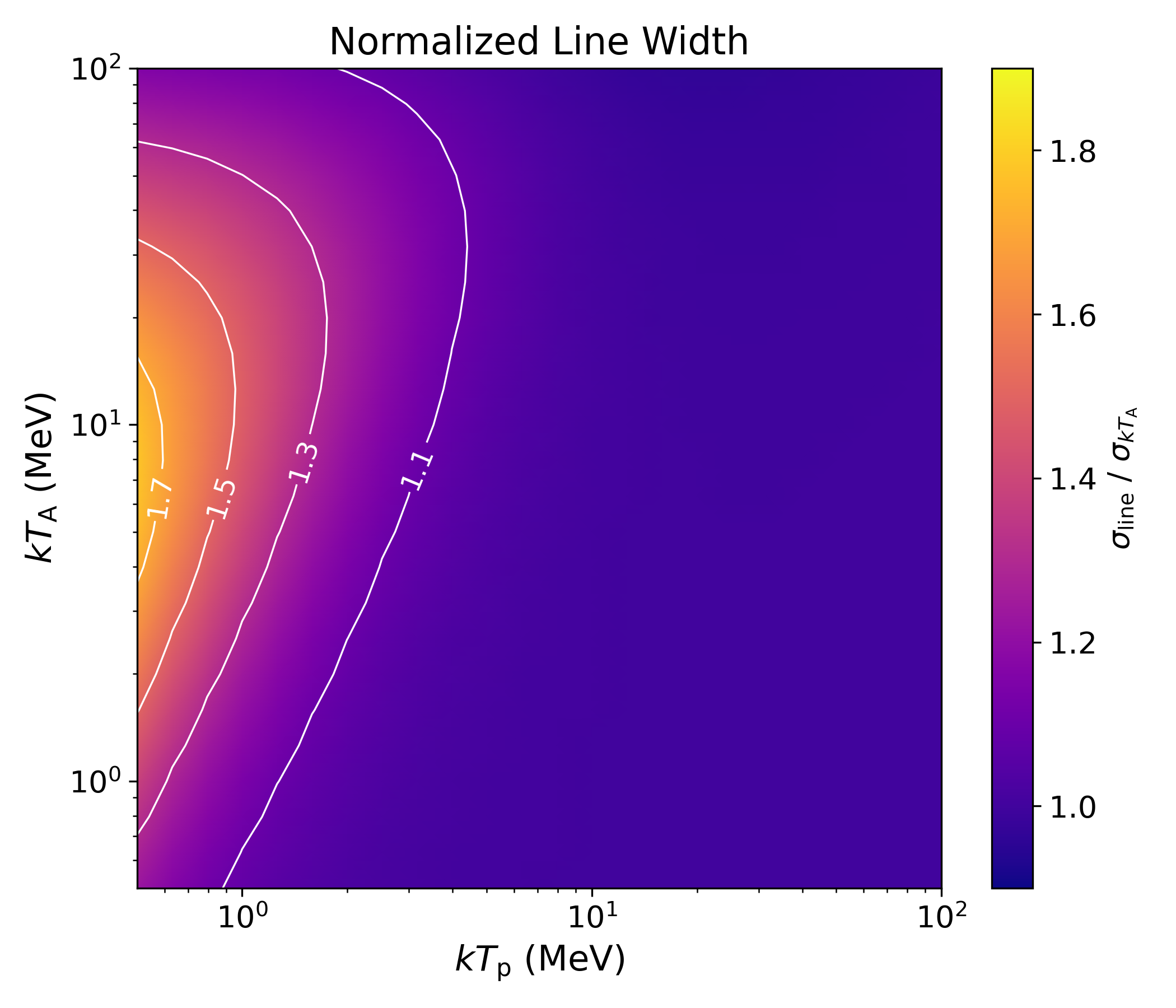}
    \caption{The gamma-ray emissivity of 4.44 MeV from $^{12}$C (p, p$\gamma$) $^{12}$C (top), line width (middle), and the normalized line width with different proton and nucleus temperatures (bottom).}
    \label{fig:param_scan_carbon}
    \end{center}
\end{figure}

Figure~\ref{fig:param_scan_carbon} shows $Q_{\gamma}$, $\sigma_{\mathrm{line}}$, and $\sigma_{\mathrm{line}} / \sigma_{kT_{\mathrm{A}}}$ for different $kT_{\mathrm{p}}$ and $kT_{\mathrm{A}}$. Here we can derive two interesting properties of the line profile. One is that the gamma-ray emissivity and line width have apparently different dependence on $kT_{\mathrm{p}}$ and $kT_{\mathrm{A}}$. The gamma-ray emissivity is mainly determined by the proton temperature. It is because to excite a nucleus by its own kinetic energy, it requires larger energy by a factor of its mass number than excitation by a proton. Especially for carbon, the nucleus kinetic energy needs to be more than 100 MeV to achieve the largest cross section at its rest frame. On the other hand, the line width is dominantly determined by the nucleus temperature, and it is less affected by the proton temperature. This is because the Doppler shift of the gamma-ray energy is determined by the nucleus motion. In principle, by combining $Q_{\gamma}$ and $\sigma_{\mathrm{line}}$ together, we can derive the temperatures for both the proton and nucleus observationally.

The other is the condition in which the line splitting is significant. As shown at the bottom of Figure~\ref{fig:param_scan_carbon}, the normalized line width is larger than one when $kT_{\mathrm{p}}$ is less than a few MeV, and it has a maximum at $kT_{\mathrm{A}} \sim 10$ MeV. This means that the line-splitting effect can be significant only with a two-temperature plasma in which $kT_{\mathrm{A}} \gg kT_{\mathrm{p}}$ and $kT_{\mathrm{p}}$ is less than a few MeV. Thus, if the line profile deviating significantly from a Gaussian profile is observed, it would be evidence that gamma rays are produced in such a two-temperature plasma.

We note that the peak height at the low-energy side of the line profile is higher than the high-energy side (see Figure~\ref{fig:lineshape_kTA_carbon}). It is because of a slightly longer tail at the high-energy side caused by the relativistic effect. When a gamma ray with the energy of $E_\gamma$ is emitted from a nucleus with a velocity of $\beta$, the maximum red-/blue-shifts will be $(\sqrt{(1 + \beta)/(1 - \beta)} - 1) \times E_\gamma$ and $(1 - \sqrt{(1 - \beta)/(1 + \beta)}) \times E_\gamma$, respectively, with the averaged energy of $E_\gamma / \sqrt{1 - \beta^2}$. Thus, as $\beta$ increases, the line becomes broader and shifts slightly towards the high-energy side. When one accumulates the line profiles produced by nuclei with different velocities in a Maxwellian distribution, the profile has a slightly longer energy tail at the high-energy side, resulting in a smaller peak height at this side.

\section{Discussion}
\label{sec_discussion}
In this work, we calculated the line profiles of gamma rays due to nuclei-proton collision. The interesting result is that the line is split when the temperature of the nuclei is much higher than that of the proton. This indicates that the line profile of the prompt gamma rays can be a smoking gun for MeV particles that are not fully thermalized in astrophysical systems. 

\subsection{Accreting black holes}
One of the potential targets is the two-temperature accretion flows onto a compact star or a supermassive black hole. It is widely accepted that at low mass accretion rates, the accretion flows become hot and optically thin because the energy advection is dominant and the radiation is inefficient \citep{Shapiro:1976,Ichimaru:1977,Ree:1982,Narayan:1994,Abramowicz:1995}.\footnote{Even at high mass accretion rates (very high state), a hot but optically-thick corona is considered to be formed \citep[e.g.,][]{Kubota2004}. If the protons and nuclei are heated to the MeV range, this state could also be a source of the de-excitation gamma-ray lines.} In contrast to the standard accretion disk model \citep{Shakura:1973}, the electrons and the protons have different temperatures in this state because electrons radiate more efficiently than protons. Also, Coulomb coupling between electrons and protons is weak due to low densities. The solution is known as the advection-dominated accretion flows \citep{Narayan:1994,Narayan1995a,Narayan1995b}. As a result, the proton temperature is close to the virial temperature in the hot accretion flow solutions, reaching $\sim 100~\mathrm{MeV} \times (r/r_s)^{-1}$, where $r$ is the distance from the central source and $r_s$ is the Schwarzschild radius (for a review, see \citealt{Yuan:2014}).

For the same reason, the protons and nuclei can have different temperatures. For example, assuming the virial temperature, the free-fall timescale becomes shorter than the proton-carbon thermal relaxation timescale around $r/r_s = 10 - 100$ when one assumes 0.01 Eddington luminosity \citep{Spitzer1962,Stepney1983}. Considering that the virial temperature is proportional to the particle mass, the temperature of the nuclei can be higher than that of the protons by a factor of their mass numbers. Furthermore, recent observations suggest the particle acceleration in the black hole corona \citep{Inoue2019}, and there is also a possibility that the particle energy distribution in the corona deviates from a Maxwellian distribution. 

\begin{figure}
    \begin{center}
    \includegraphics[width=8.0 cm]{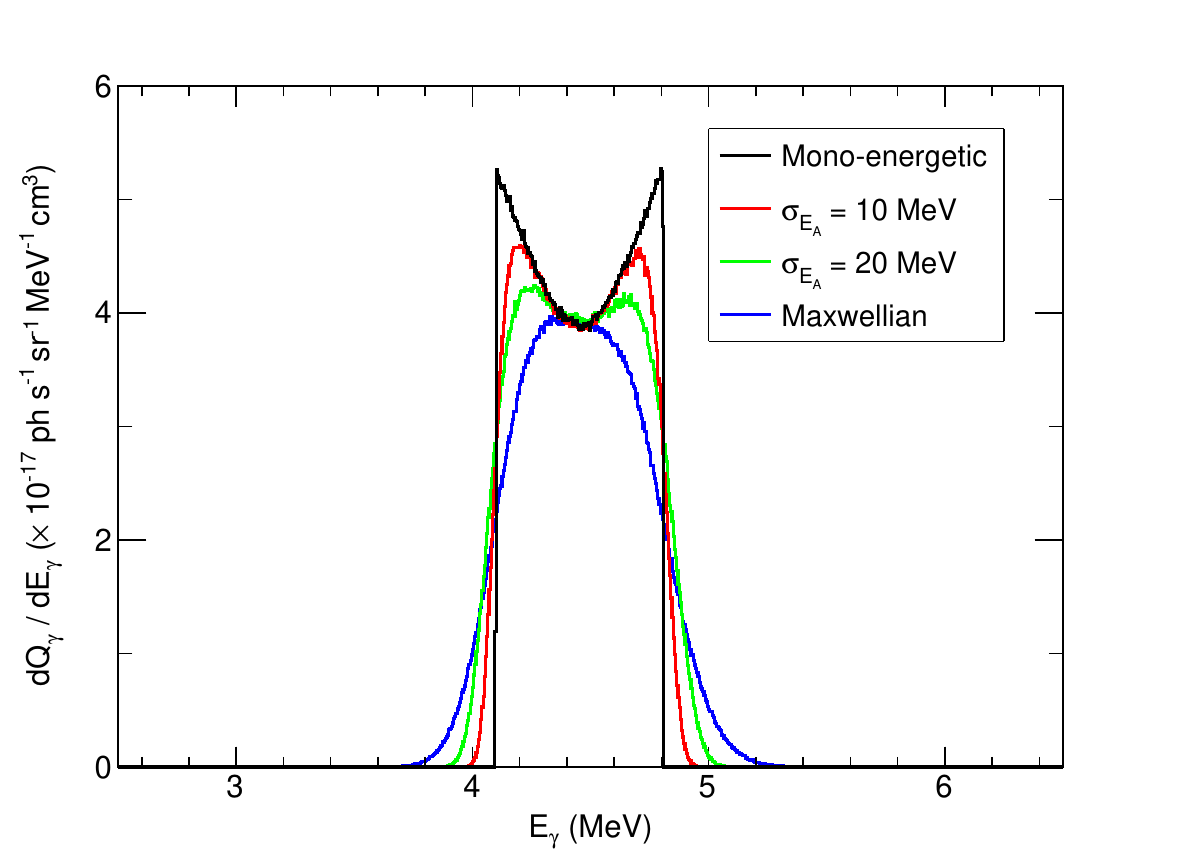}
    \includegraphics[width=8.0 cm]{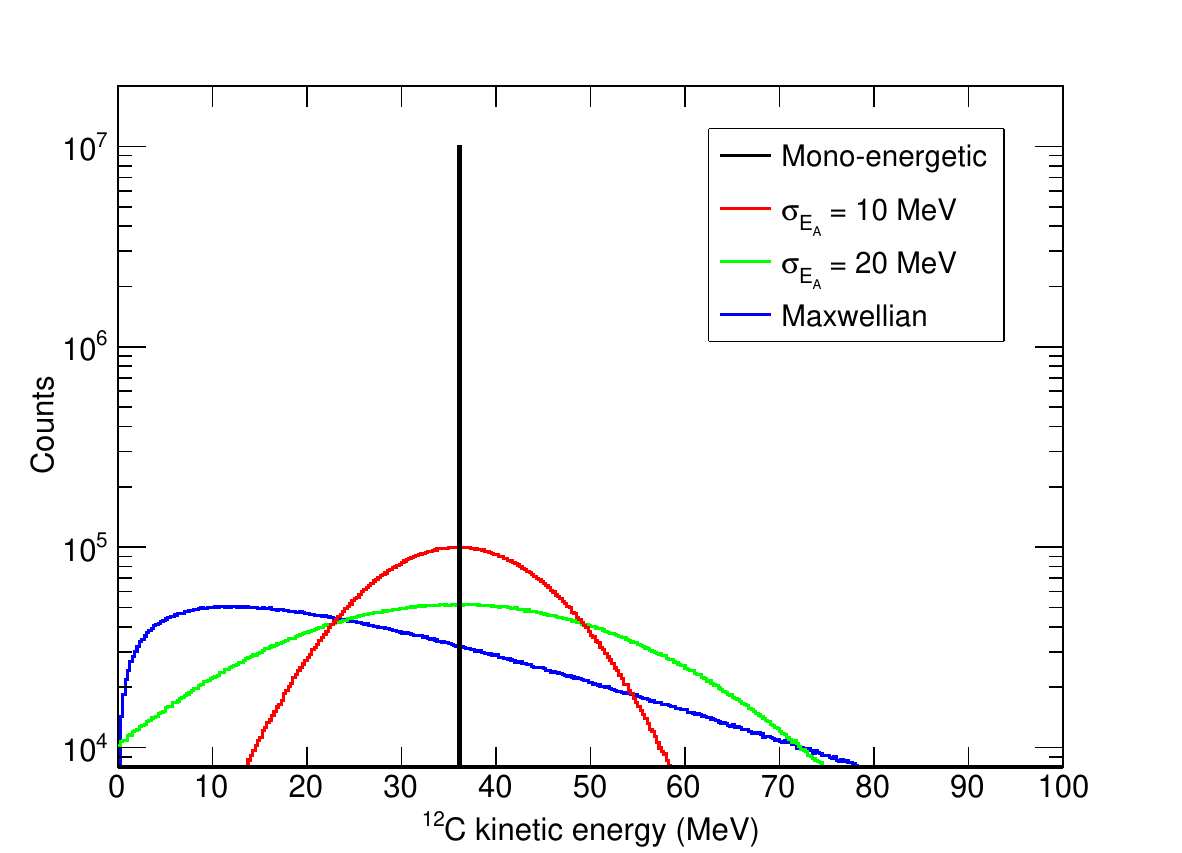}
    \caption{4.44 MeV gamma-ray line profiles from different nucleus energy distributions (top). The bottom panel shows the energy distribution of each case. Here the averaged kinetic energy of Carbon is the same as 36 MeV over all cases. The proton is assumed to be Mawellian with $kT_{\mathrm{p}}$ of 2 MeV. The black line corresponds to the mono-energetic case. The red and green lines are the results of Gaussian energy distributions with $\sigma_{E_k}$ of $10~\mathrm{MeV}$ and $20~\mathrm{MeV}$, respectively. In the blue lines, the Maxwellian distribution is assumed.}
    \label{fig:kTA24_kTp2}
    \end{center}
\end{figure}

In order to demonstrate how the line profile depends on the particle energy distribution and the temperatures, here we assume several nuclei energy distributions and temperatures and outline some examples. Considering the hot accretion flow and the virial temperature, we set the average kinetic energy of nuclei to twelve times the kinetic energy of protons, as the gravitational energy release is proportional to the particle mass. The proton energy distribution is assumed to be a Maxwellian distribution.
Figure~\ref{fig:kTA24_kTp2} shows the line profile with $kT_\mathrm{p} = 2~\mathrm{MeV}$.
Since the average energy as a function of temperature $T$ is $\frac{3}{2}kT$, the nucleus kinetic energy is set to $36~\mathrm{MeV} (= 1.5 \times 12\times2.0~\mathrm{MeV})$. The red and green lines are the results with Gaussian energy distributions with a standard deviation ($\sigma_{E_{\mathrm{A}}}$) of 10 and 20 MeV, respectively. The blue line is the case when the energy distribution is Maxwellian. 
We note that the nucleus energy distributions in Figure~\ref{fig:kTA24_kTp2} have the same average energy, and these distributions trace the relaxation process from mono-energetic to thermal equilibrium. Our calculations show that we can disentangle the particles' energy distributions by detailed observations of nuclear line profiles.

For a more detailed discussion, we need to account for the thermal distribution depending on the radial axis, other nuclear processes (see Section~\ref{sec_csdata}), and gravitational Doppler shift etc., which is a subject for future work. Especially the radial dependence is critical to discuss the detectability of this effect because the observable line profile is the superposition of the lines at different radii. Since the protons and nuclei vary their temperatures with radius, the line-splitting effect could be suppressed and challenging to observe. However, the line gamma rays may be produced dominantly in the outer range of the corona (100--1000 $r_s$) if the nuclear spallation reduces carbon's fraction significantly as it reaches the center (see Figure 3 in \citealt{Ervin:2018b}). Since the proton temperature is a few MeV at these radii, the line may still deviate from Gaussian in such cases. Also, the line profiles may be partially smeared by scattering processes very close to the central source because the Thomson optical depth is about $\sim 1 \times (r/r_s)^{-0.5}$ with $10^{-2}$ Eddington luminosity. Here we adopt the solution shown in \cite{Narayan:1994} with $\alpha c_1 = 0.05$ and $c_3 = 0.5$. We should also note that the continuum emission from the corona, e.g., the inverse Compton scattering and thermal electron Bremsstrahlung \citep{Ervin:2018b,Inoue2019}, may overwhelm the line emission, causing less detectability.

We finally mention the line gamma-ray observations with future MeV gamma-ray satellites. Based on the spectral models by \citet{Ervin:2018a,Ervin:2018b}, the 4.44 MeV gamma-ray flux from a source at the distance of 2 kpc, such as Cygnus X-1, is estimated as $10^{-9}$ to $10^{-6}~\mathrm{ph/cm^{2}/s}$ depending on the assumed disk model and its parameters. 
The upcoming MeV gamma-ray satellite, the Compton Spectrometer and Imager will have a line sensitivity of a few $\times 10^{-6}~\mathrm{ph/cm^{2}/s}$ at 4.4 MeV \citep{COSIofficial}.
While it is not likely to observe the 4.4 MeV line, we would note that the upper bound of these fluxes is within reach if the observations are longer than the proposed and accepted two-year mission.
Future MeV gamma-ray observations, e.g., AMEGO \citep{AMEGO2019} and GRAMS \citep{ARAMAKI2020107} are expected to have a better line sensitivity at 4.4 MeV, so that the brightest cases might be reached quicker. 

\subsection{Supernova remnants}

The de-excitation gamma-ray lines are also used to infer the conditions of particle acceleration operated in supernova remnants \citep{Liu2021} and solar flares \citep{Kiener2006,Smith2003}. One of the delectable targets is Cassiopeia A. \cite{Summa2011} modeled the nuclear de-excitation line spectrum from Cassiopeia A, predicting that the 4.44 and 6.13 MeV lines are close to the sensitivity limit of COMPTEL. If the de-excitation lines are detected with future MeV missions, it can give information about low-energy cosmic rays, such as injection rate, gas density, and energy spectrum. As introduced in Section~\ref{sec_intro}, the line-splitting effect can also occur in the non-thermal case if the accelerated nuclei collide with ambient protons \citep{Bykov:1996}. In addition, there is a possibility that the two-temperature plasma is produced via the mass-proportional heating in the collisionless shock \citep{Rakowski:2005,Ghavamian:2013} though it is not clear whether it dominates over the non-thermal particles. Considering that the line profile depends on the particle energy distributions, as shown in Figure~\ref{fig:kTA24_kTp2}, if the de-excitation lines are observed from supernova remnants, their line profile might be key observables to investigate the energy distribution of the produced MeV particles and the energy injection problem of low-energy cosmic rays.

\section{Conclusion}
\label{sec_conclusion}
In this paper, we have calculated the line profile of prompt gamma rays from $^{12}$C and $^{16}$O as examples considering the anisotropic gamma-ray emission in the nucleus rest frame. We found that the line splitting can take place in a hot two-temperature plasma in which the proton temperature is less than a few MeV, and the nucleus temperature is high enough to produce nuclear de-excitation gamma rays. In contrast, the gamma-ray line has a Gaussian-like shape when the proton temperature becomes higher and close to the nucleus temperature. Our results suggest that the profile of prompt gamma-ray lines can be a tool to probe for physical conditions in a very hot thermal plasma. It can be applied to investigate the particle distribution of nuclei and protons in the hot accretion flow around black holes, where protons and nuclei can have different temperatures of $>$ few MeV. 
In addition, the particle injection of the acceleration process in supernova remnants might be studied by observations of gamma-ray line splitting. 
Considering the width of the split lines shown in Figure~\ref{fig:lineshape_kTA_carbon} and \ref{fig:lineshape_kTA_oxygen} are 8\% and 2\% (1 $\sigma$), respectively, future MeV gamma-ray observations with an energy resolution of a few \% at least are needed to enable such a new plasma diagnostics. 

\section{Data Availability}
The cross-section data of the nuclear reactions used in this work are available in the articles listed in Section~\ref{sec_csdata}. The code developed for the line profile calculation in this work will be shared on reasonable request to the corresponding author.
  
\section*{Acknowledgements}
The authors would like to thank the anonymous referee, Karl Mannheim and Dmitry Khangulyan for useful comments.
Hiroki Yoneda acknowledges support by the Bundesministerium f\"{u}r Wirtschaft und Energie via the Deutsches Zentrum f\"{u}r Luft- und Raumfahrt (DLR) under contract number 50 OO 2219. 
Tadayuki Takahashi acknowledges support from JSPS KAKENHI grant number 20H00153.




\input{output.bbl}



\appendix

\section{6.13 MeV gamma-ray from \texorpdfstring{$^{16}$}{16}O-p collision}
\label{sec:oxygen}

\begin{figure}
    \begin{center}
    \includegraphics[width=8.0 cm]{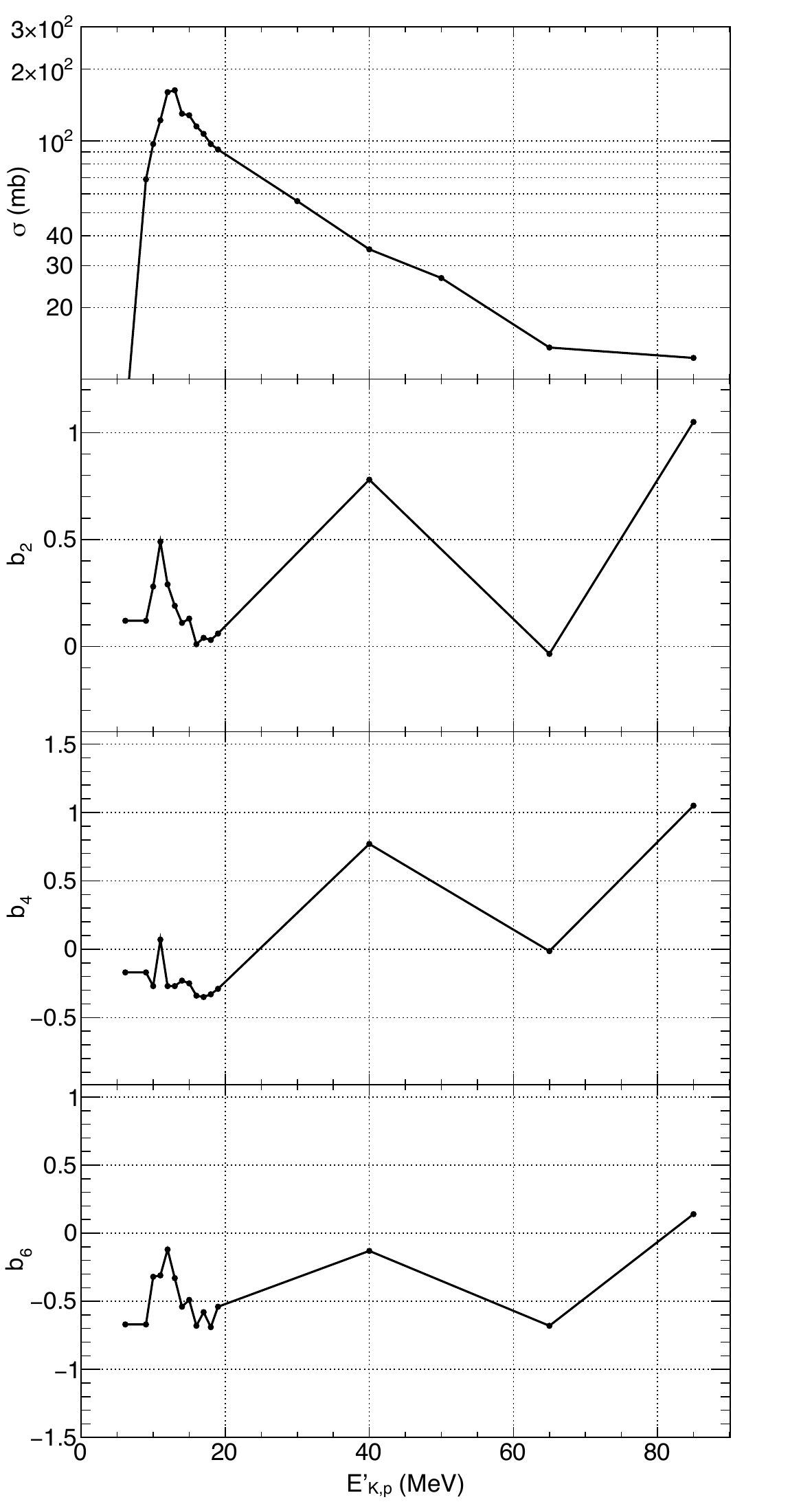}
    \includegraphics[width=8.0 cm]{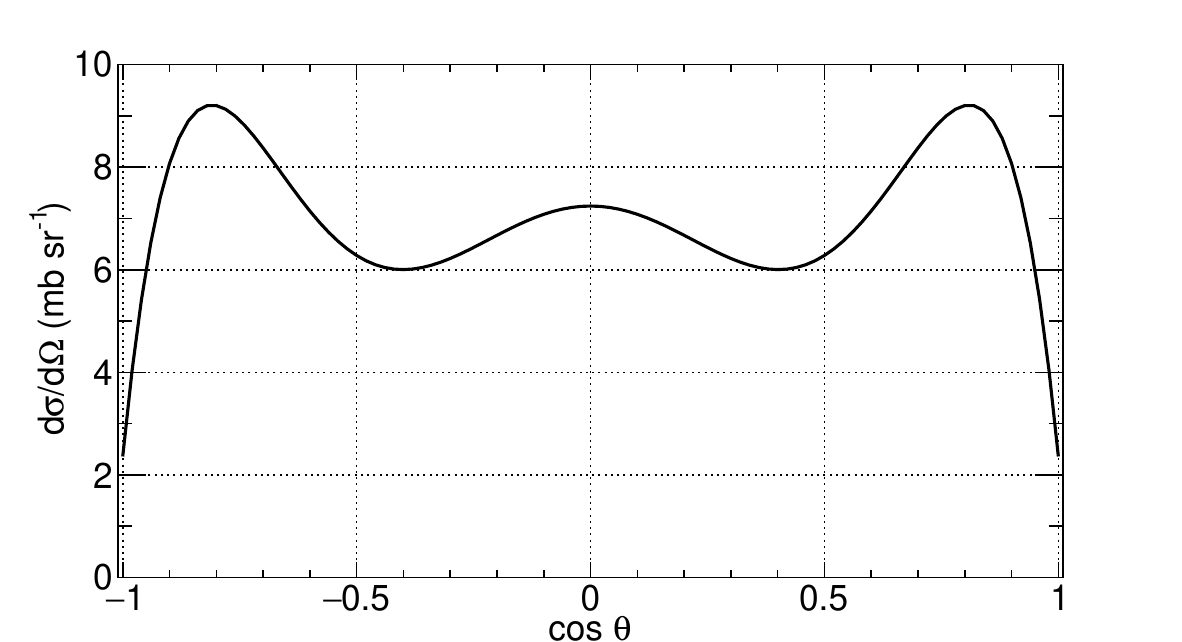}
    \caption{The cross section data used for 6.13 MeV gamma-ray emission from $^{16}$O (p, p$\gamma$) $^{16}$O. The top is the total cross section. The second, third, and fourth are the coefficient $b_2$, $b_4$, and $b_6$, respectively. The bottom is the differential cross section at $E'_\mathrm{K,p} = 20$ MeV, shown as an example.}
        \label{fig:cs_oxygen}
    \end{center}
\end{figure}

\begin{figure}
\begin{center}
    \includegraphics[width=8.0 cm]{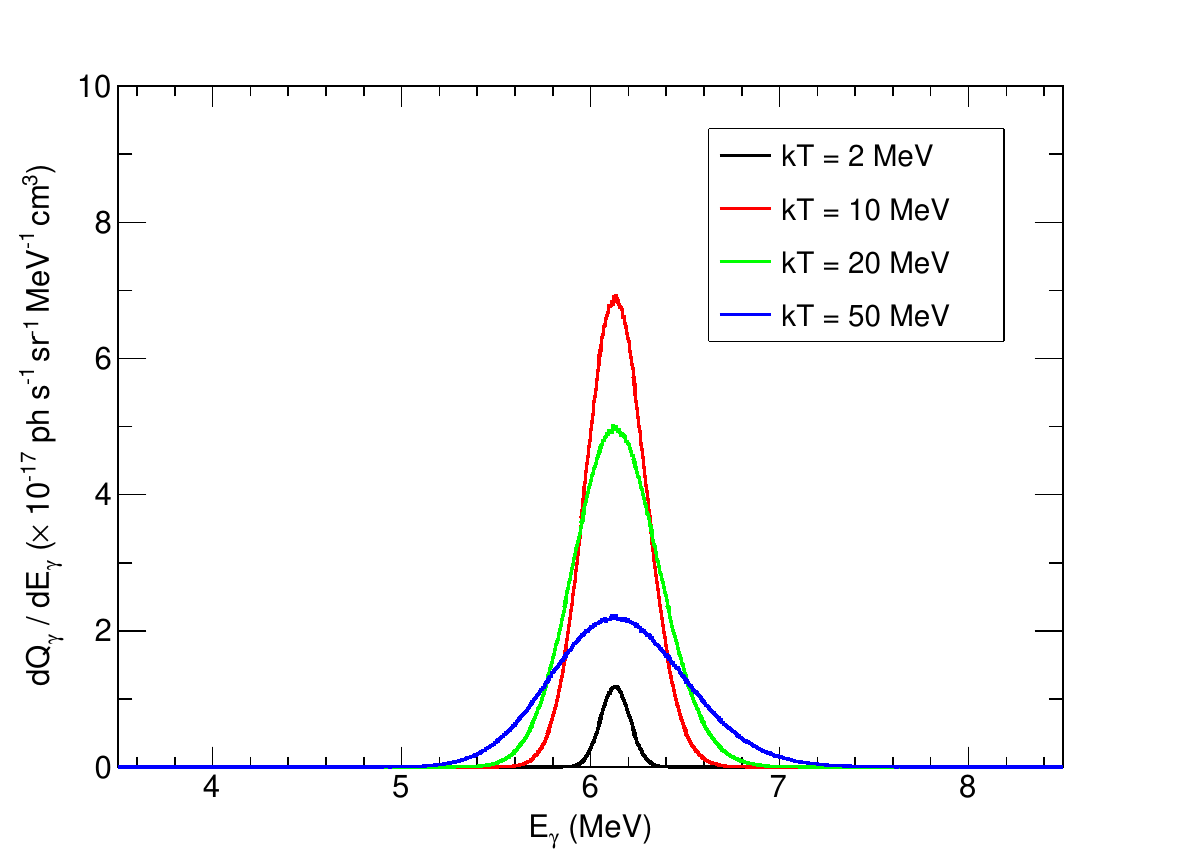}
    \caption{6.13 MeV gamma-ray line profile assuming $T_\mathrm{p} = T_\mathrm{A}$. Black: $kT_\mathrm{A} = kT_\mathrm{p} = 2.0~\mathrm{MeV}$, Red: $10.0~\mathrm{MeV}$, Green: $20.0~\mathrm{MeV}$, Blue: $50.0~\mathrm{MeV}$.}
\label{fig_oxygen_sameTemp}
\end{center}
\end{figure}

\begin{figure}
    \begin{center}
    \includegraphics[width=8.0 cm]{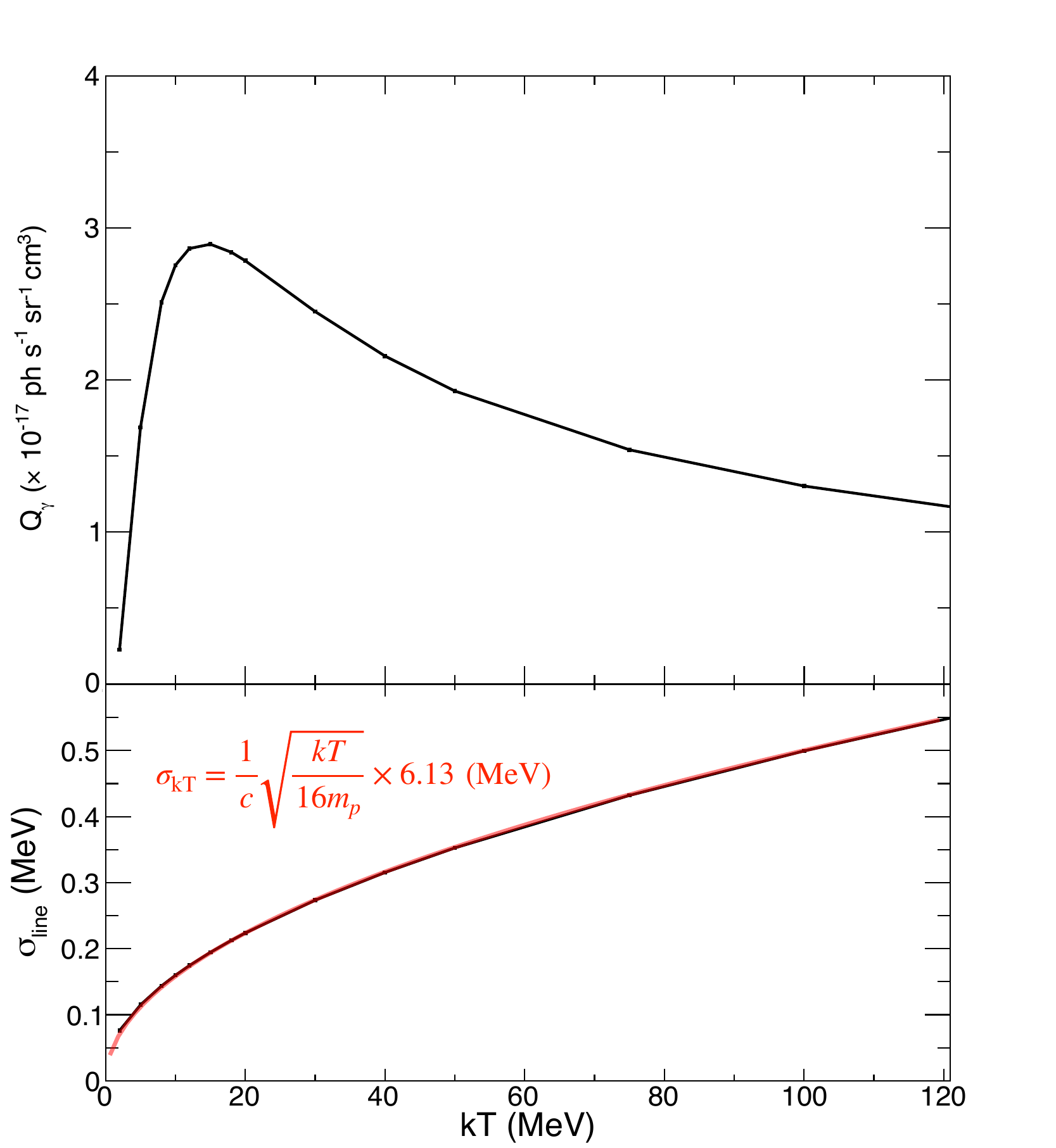}
    \caption{The dependence of the gamma-ray emissivity and the line width of 6.13 MeV gamma ray from $^{16}$O on the temperature, assuming $T_\mathrm{p} = T_\mathrm{A}$. The top panel shows the gamma-ray emissivity depending on the temperature, and the bottom is the standard deviation of the line profile. The red line is the value expected from the normal thermal Doppler broadening.}
    \label{fig_oxygen_sameTemp_flux_sigma}
    \end{center}
\end{figure}

\begin{figure}
    \begin{center}
    \includegraphics[width=8.0 cm]{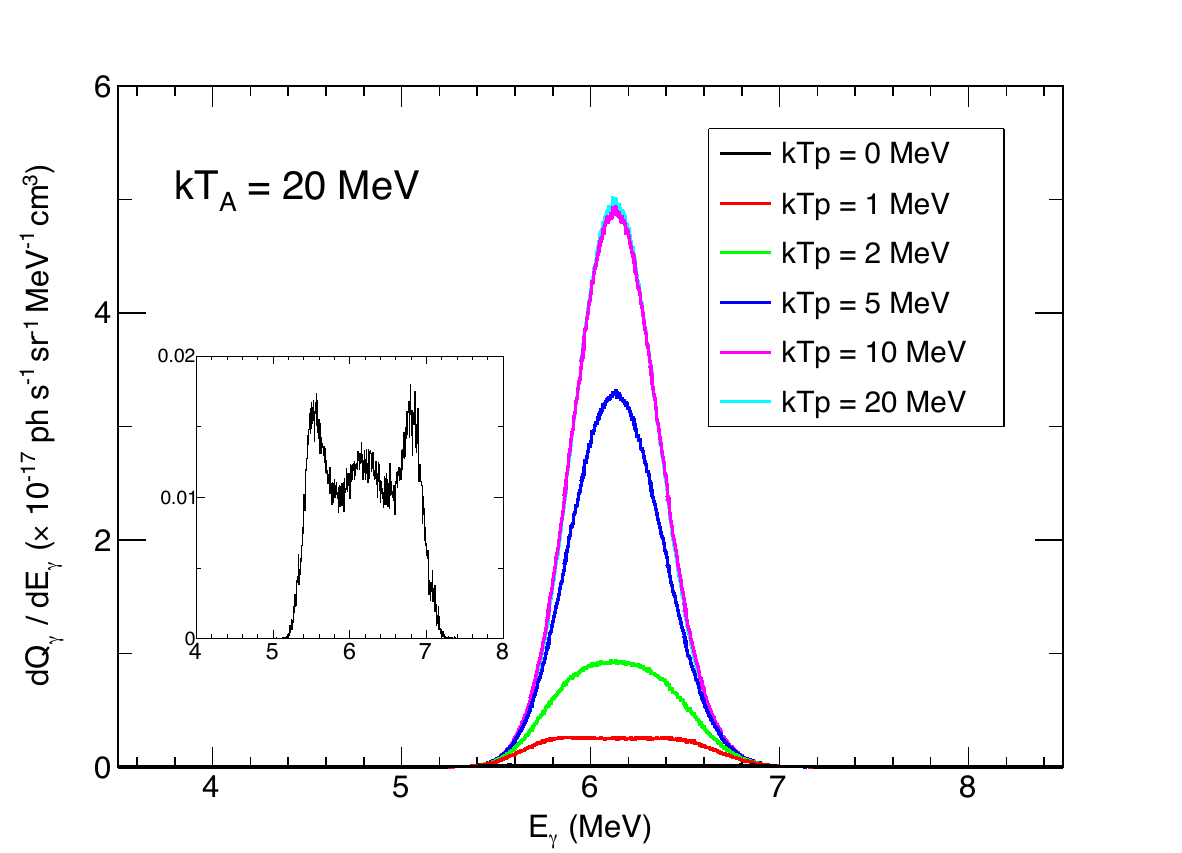}
    \includegraphics[width=8.0 cm]{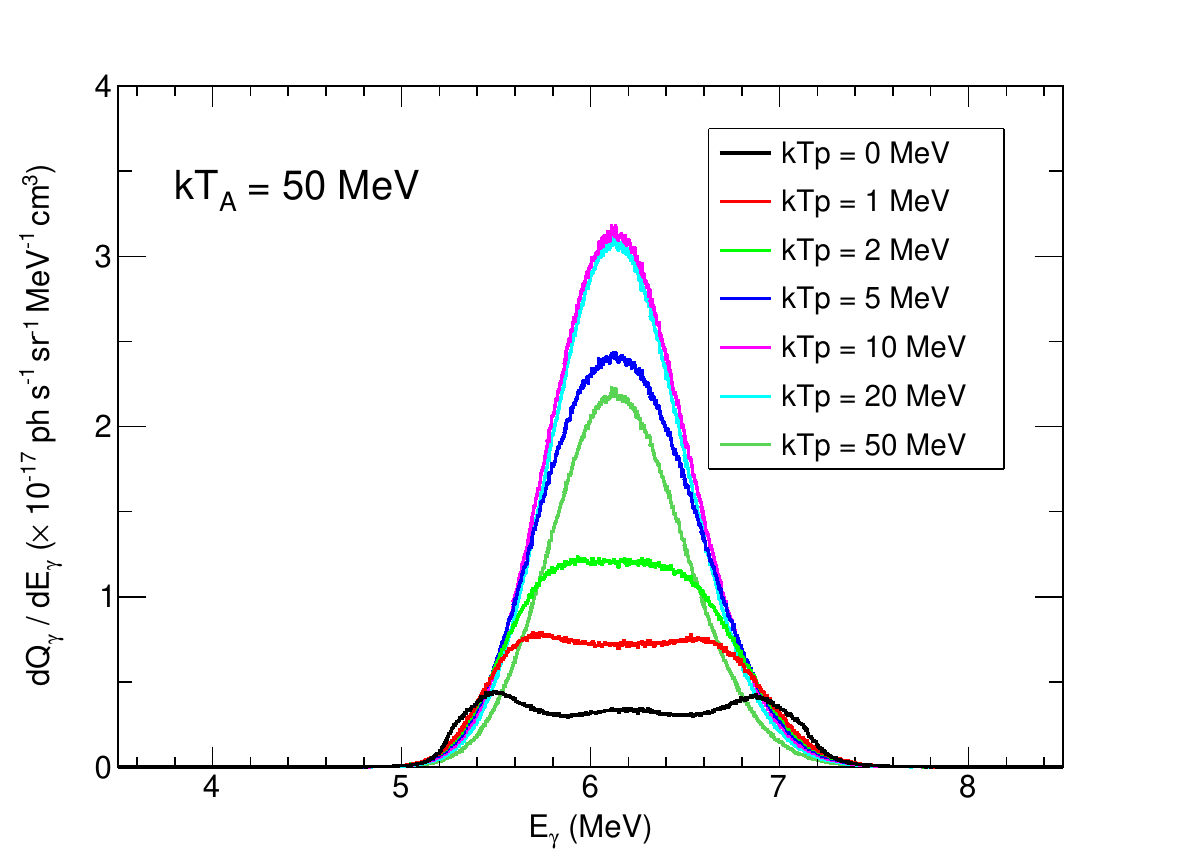}
    \caption{The line profile of 6.13 MeV gamma-ray line with different proton temperatures. The nucleus temperature is fixed to 20 MeV (top) or 50 MeV (bottom). The inset in the top panel is a zoom of the line profile with $kT_\mathrm{p}$ = 0 MeV.}
    \label{fig:lineshape_kTA_oxygen}
    \end{center}
\end{figure}

\begin{figure}
    \begin{center}
    \includegraphics[width=8.0 cm]{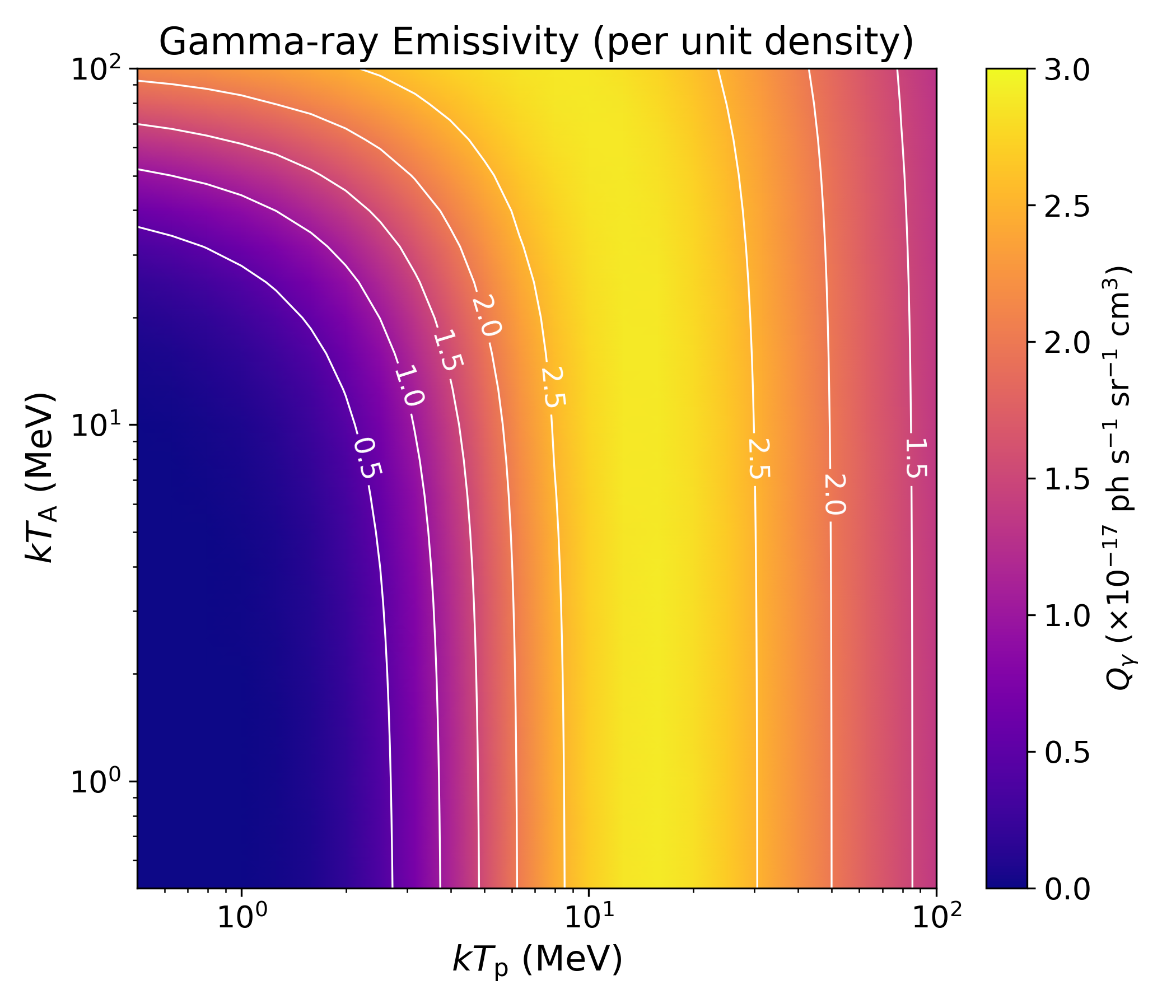}
    \includegraphics[width=8.0 cm]{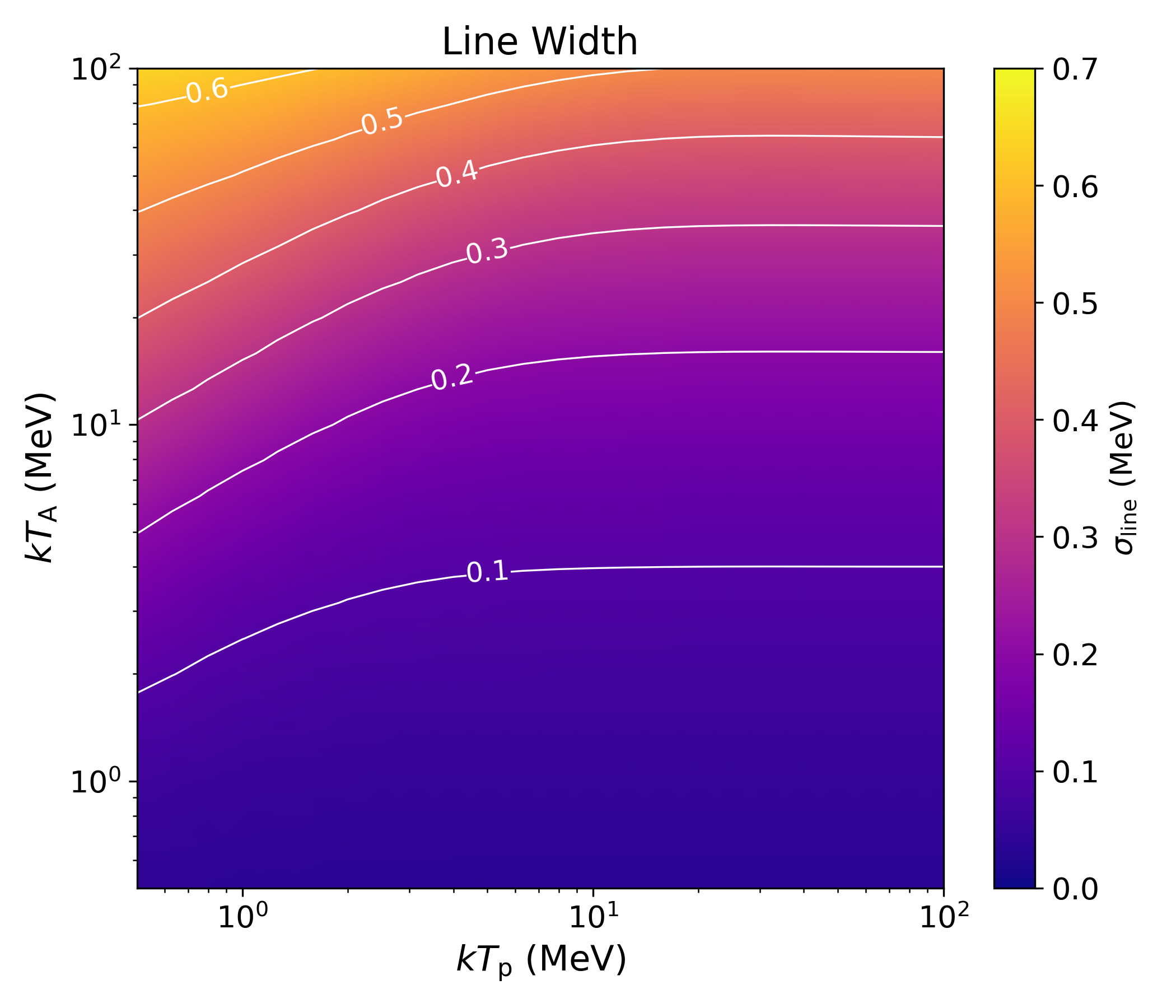}
    \includegraphics[width=8.0 cm]{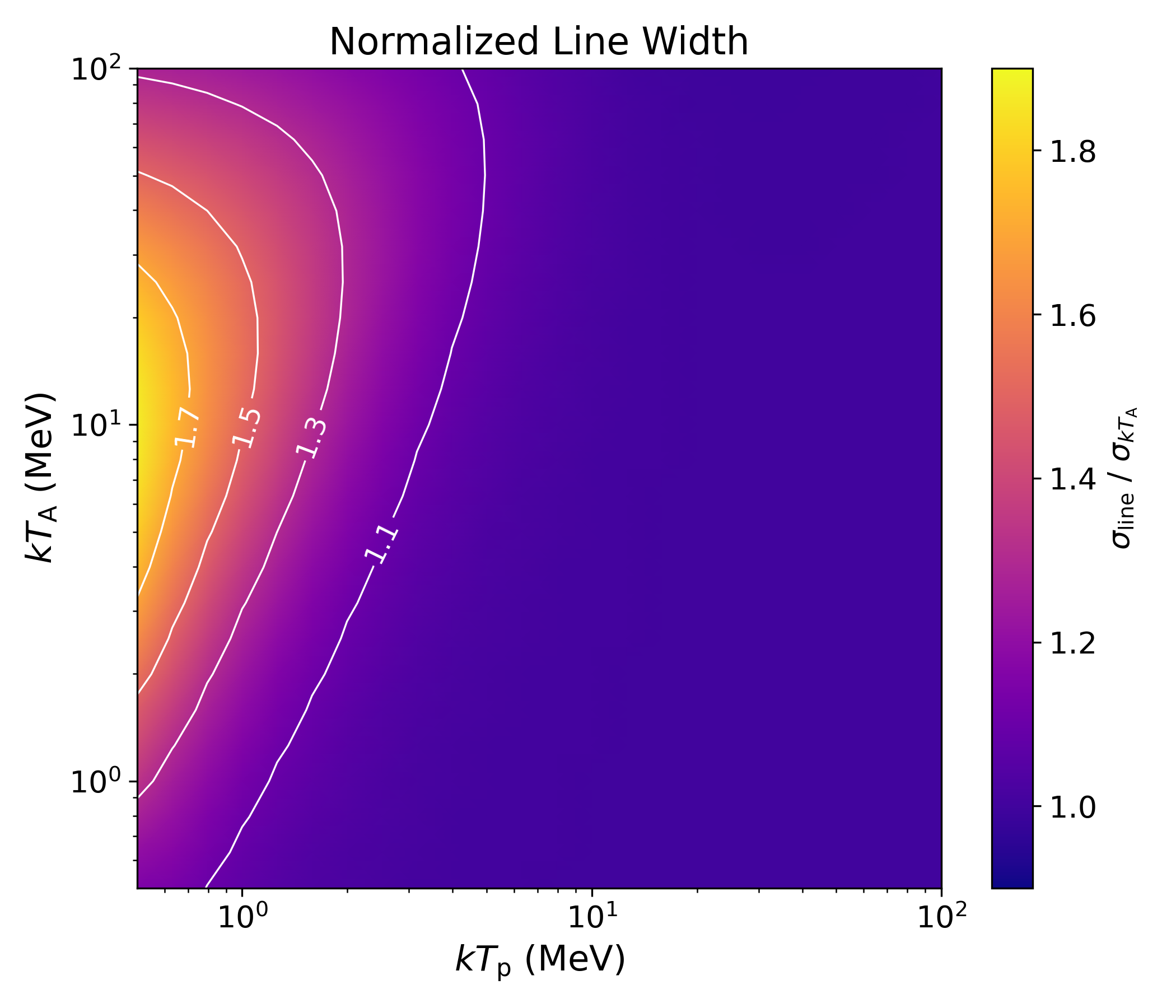}
    \caption{The same as Figure~\ref{fig:param_scan_carbon}, but for 6.13 MeV from $^{16}$C (p, p$\gamma$) $^{16}$O.}
    \label{fig:param_scan_oxygen}
    \end{center}
\end{figure}

Using the cross-section dataset shown in Figure~\ref{fig:cs_oxygen}, We show our results in the case of $^{16}$O in Figures ~\ref{fig_oxygen_sameTemp}, ~\ref{fig_oxygen_sameTemp_flux_sigma}, ~\ref{fig:lineshape_kTA_oxygen}, ~\ref{fig:param_scan_oxygen}. When the two temperatures are the same, the line has a Gaussian-like shape (Figure~\ref{fig_oxygen_sameTemp}), and the gamma-ray emissivity traces the cross section (Figure~\ref{fig_oxygen_sameTemp_flux_sigma}). These results are the same as those in the case of carbon. A difference is the line profile when the line splitting occurs. As shown in Figure~\ref{fig:lineshape_kTA_oxygen}, the gamma-ray line can have three peaks because the differential cross section has three peaks. Figure~\ref{fig:cs_oxygen} shows the differential cross section at $E = 20$ MeV. We observe the line splitting only when the proton energy is lower than a few MeV, as is the case of carbon (see the bottom panel of Figure~\ref{fig:param_scan_oxygen}).

\section{The effect of recoil energy in the collision}
\label{sec_recoil}
\cite{Kozlovsky:1997} calculate the line profile, including the recoil energy in the case that the nuclei are at rest and protons have an energy of 23 MeV. In their approach, it is assumed that the anisotropic gamma-ray emission depends on $\theta_{w}$,  the angle between the direction of the gamma ray and the normal to the reaction plane in the center-of-mass frame. However, it is difficult to apply this approach to our calculation because there is no data to describe the anisotropic gamma-ray emission with $\theta_{w}$ in the energy range up to $\sim$ 100 MeV. They also consider that the recoiling nuclei are moving preferentially backward in the center-of-mass frame \citep{ramaty_nuclear_1979}, but this data is also lacking in the wide energy range.

Alternatively, we calculate the gamma-ray energy in the nucleus rest frame instead of fixing it as follows. In advance of the Monte Carlo calculation in Section~\ref{sec_calc}, we calculate the gamma-ray energy distribution in the nucleus rest frame under assumptions that (1) the recoiling nuclei move isotropically in the center-of-mass frame and (2) the gamma-ray emission is isotropic in the rest frame of an excited nucleus. We sample the excited nuclei and gamma-ray momenta and convert the gamma-ray energy into the nucleus rest frame. By repeating it, we obtain the gamma-ray energy distribution in the nucleus rest frame depending on the angle between the incoming proton and the gamma-ray direction. Then, at step (v) of the Monte Carlo calculation, we sample the gamma-ray energy in the nucleus rest frame from the pre-calculated gamma-ray energy distribution.

\begin{figure}
\begin{center}
\includegraphics[width = 8.0 cm]{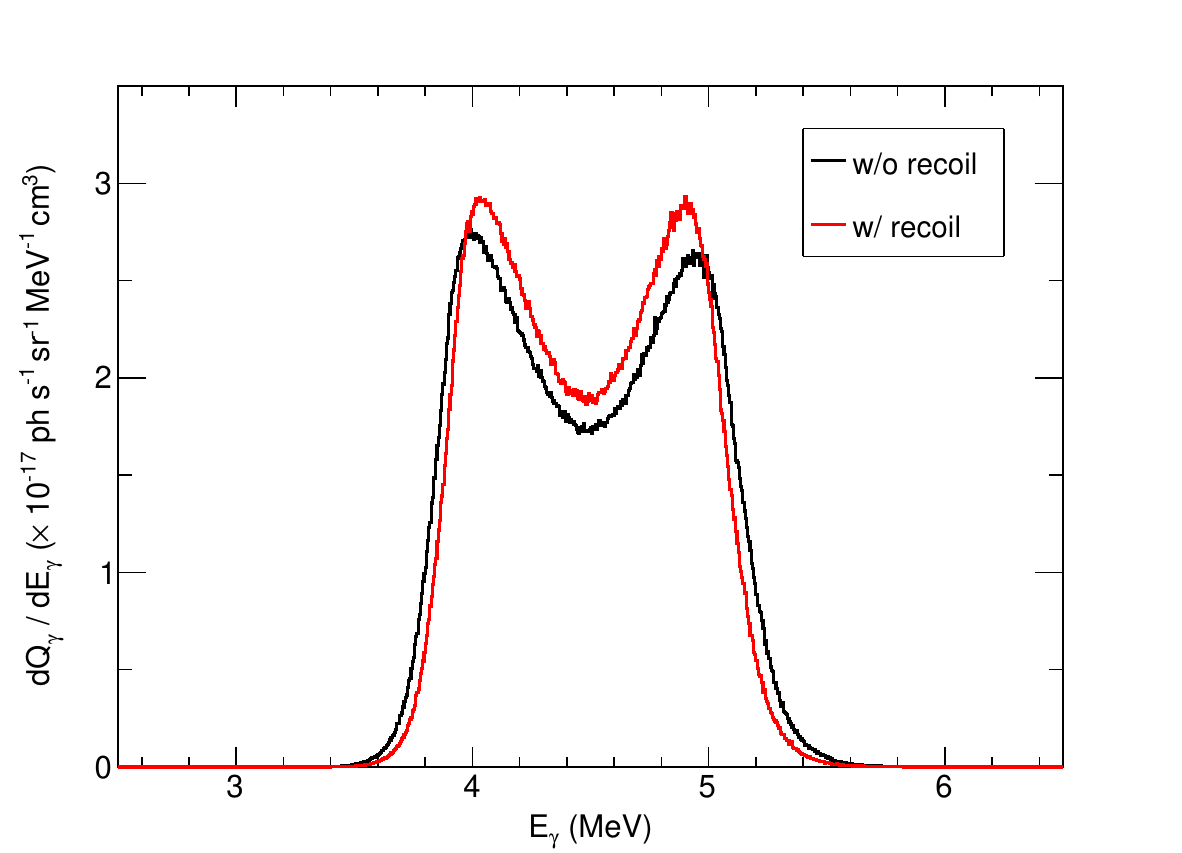}
\end{center}
\caption{4.44 MeV line profile considering the effect of the recoil energy. The black and red lines correspond to the spectra without and with the recoil effect. Here $kT_{\mathrm{p}}$ and $kT_{\mathrm{A}}$ are set to 0 and 50 MeV, respectively.}
\label{fig_recoil_thermal}
\end{figure}

Figure~\ref{fig_recoil_thermal} shows an example of the line profile following the calculation considering the recoil effect. We can see that the recoil effect makes the line profile slightly narrower because the nuclei lose their kinetic energy by the excitation energy, and then the Doppler shift becomes smaller. Thus, the peak in the spectrum with the effect is higher than that without the effect. The difference of the peak height is $\sim 5\%$ in Figure~\ref{fig_recoil_thermal}, where we assume that $kT_{\mathrm{p}} = 0$ MeV and $kT_{\mathrm{A}} = 50$ MeV. Though the recoil effect changes the line shape slightly, the line-splitting feature is kept even after including the effect. We should note that assumption (2) in the pre-calculation is inconsistent with the actual anisotropic gamma-ray distribution. To improve this point, we would need more experimental data about $^{12}$C (p, p$\gamma$) $^{12}$C, including the recoiling effect.

\section{
Validity of the approximation of the cross section in low energy range.}
\label{sec:approximation_validity}

\begin{figure}
    \begin{center}
    \includegraphics[width=8.0 cm]{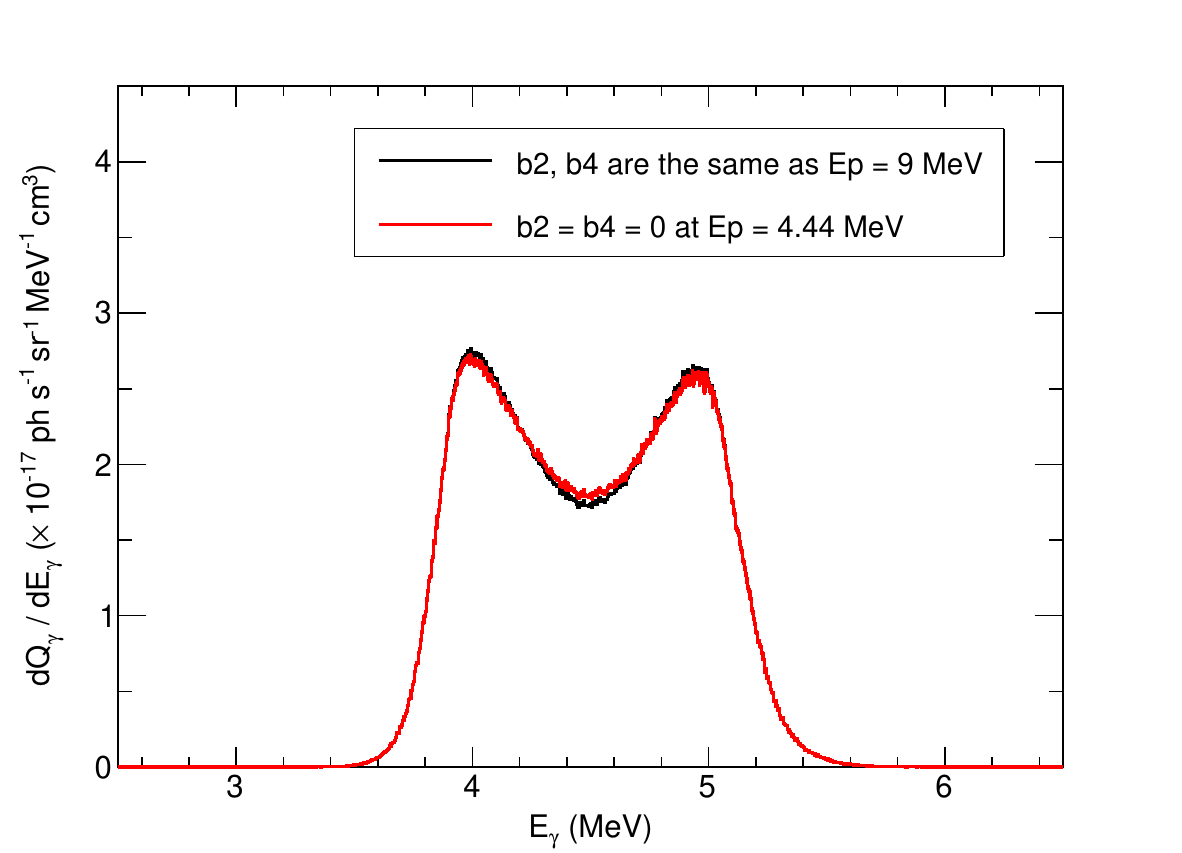}
    \caption{Comparison between the different approximations of the cross section in low energy range.
    Black: $b_2, b_4$ at 4.44 MeV are the same as those at 9 MeV,
    Red: $b_2, b_4$ are zeros at 4.44 MeV.
    Here $kT_{\mathrm{A}} = 50.0~\mathrm{MeV}$ and $kT_{\mathrm{p}}=0.0~\mathrm{MeV}$ are adopted.}
    \label{fig:comp}
        \end{center}
\end{figure}

For $^{12}$C (p, p$\gamma$) $^{12}$C, due to the lack of experimental data, we assumed that the coefficients $b_2, b_4$ between 4.44 and 9 MeV are the same as those measured at 9 MeV in the main text. However, we need to clarify whether this assumption is valid or not. We adopt a different approximation that the coefficients $b_2, b_4$ are zero at 4.44 MeV and compare the difference in the calculated line profiles. Note that with the new assumption, the gamma-ray emission is isotropic close to the threshold in the nuclei rest frame. Figure~\ref{fig:comp} is the result with $kT_{\mathrm{A}} = 50.0~\mathrm{MeV}$ and $kT_{\mathrm{p}}=0.0~\mathrm{MeV}$. There is a difference of a few percent around the energy center, which is trivial and does not affect our discussion.

\bsp	
\label{lastpage}
\end{document}